\documentclass[aps,jmp,amsmath,amssymb,reprint]{revtex4-1}

\usepackage{graphicx}
\usepackage{dcolumn}
\usepackage{bm}
\usepackage{amssymb}
\usepackage{amsmath}
\usepackage{subfigure}
\usepackage{physics}
\usepackage{sublabel}
\usepackage[utf8]{inputenc}

\usepackage{hyperref}
\usepackage[usenames, dvipsnames]{color}
\usepackage[english]{babel}

\usepackage{bbm}
\hypersetup{colorlinks = true, linkcolor=blue, citecolor=blue, urlcolor=}

\usepackage[T1]{fontenc}

\begin{document}


\title{Role of dephasing on the conductance signatures of Majorana zero modes
}

\author{Chaitrali Duse}
\affiliation{Department of Physics, Indian Institute of Technology Bombay, Powai, Mumbai-400076, India}
\author{Praveen Sriram}
\affiliation{Department of Applied Physics, Stanford University, 348 Via Pueblo, Stanford, CA 94305, United States of America}
\author{Kaveh Gharavi}
\affiliation{Institute for Quantum Computing, University of Waterloo, Waterloo, Ontario, Canada N2L 3G1}
\affiliation{Department of Physics and Astronomy, University of Waterloo, Waterloo, Ontario, Canada N2L 3G1}
 \author{Jonathan Baugh}
 \affiliation{Institute for Quantum Computing, University of Waterloo, Waterloo, Ontario, Canada N2L 3G1}
\affiliation{Department of Physics and Astronomy, University of Waterloo, Waterloo, Ontario, Canada N2L 3G1}
\affiliation{Department of Chemistry, University of Waterloo, Waterloo, Ontario, Canada N2L 3G1}
\author{Bhaskaran Muralidharan}
\affiliation{Department of Electrical Engineering, Indian Institute of Technology Bombay, Powai, Mumbai-400076, India}
\email{bm@ee.iitb.ac.in}

\date{\today}
\begin{abstract}
Conductance signatures that signal the presence of Majorana zero modes in a three terminal nanowire-topological superconductor hybrid system are analyzed in detail, in both the clean nanowire limit and in the presence of non-coherent dephasing interactions. In the coherent transport regime for a clean wire, we point out contributions of the local Andreev reflection and the non-local transmissions toward the total conductance lineshapes while clarifying the role of contact broadening on the Majorana conductance lineshapes at the magnetic field parity crossings. Interestingly, at larger $B$-field parity crossings, the contribution of the Andreev reflection process decreases which is compensated by the non-local processes in order to maintain the conductance quantum regardless of contact coupling strength. In the non-coherent transport regime, we include  dephasing that is introduced by momentum randomization processes, that allows one to smoothly transition to the diffusive limit. Here, as expected, we note that while the Majorana character of the zero modes is unchanged, there is a reduction in the conductance peak magnitude that scales with the strength of the impurity scattering potentials. Important distinctions between the effect of non-coherent dephasing processes and contact-induced tunnel broadenings in the coherent regime on the conductance lineshapes are elucidated. Most importantly our results reveal that the addition of dephasing in the set up does not lead to any notable length dependence to the conductance of the zero modes, contrary to what one would expect in a gradual transition to the diffusive limit. We believe this work paves a way for a systematic introduction of scattering processes into the realistic modeling of Majorana nanowire hybrid devices and assessing topological signatures in such systems in the presence of non-coherent scattering processes. 
\end{abstract}

\maketitle


\section{\label{sec:intro}Introduction}
\begin{figure}
\centering
\subfigure[]{
    \includegraphics[width=\columnwidth]{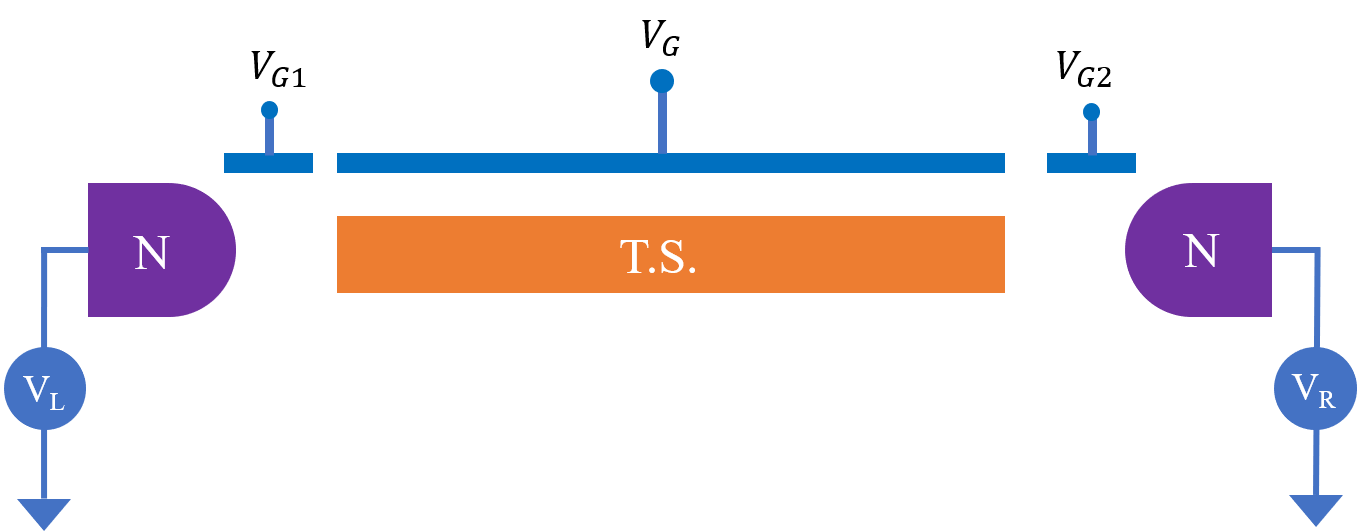}
    \label{fig:schematic-a}}
 \\
\subfigure[]{
    \includegraphics[width=\columnwidth]{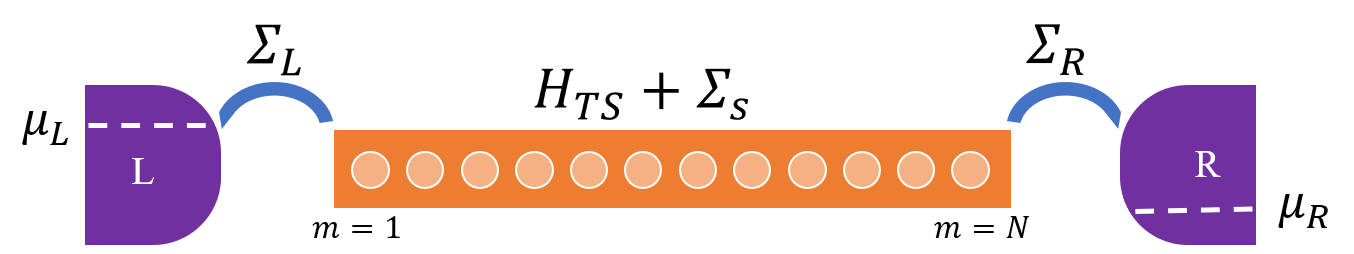}
    \label{fig:schematic-b}}
\label{fig:schematic}
\caption{ Device schematics: (a) The N-TS-N system under consideration in this work. $V_G$ controls the chemical potential in the nanowire, which we assume to be spatially uniform. $V_{G1}$ and $V_{G12}$ control the coupling to the normal contacts on either side. Voltages $V_L$ and $V_R$ are applied to the respective contacts. Throughout this work, we will be focusing on symmetric biasing ($V_L = -V_R = V/2$). (b) After discretizing the nanowire as a N-site tight-binding lattice (with site index m), the self-energy matrices $\Sigma_L$ and $\Sigma_R$ in the NEGF formalism take into account the coupling of the channel to the contacts. Non-coherent scattering processes within the nanowire are taken into account by the scattering self-energy $\Sigma_R$. }
\end{figure}
Quasi one-dimensional Rashba nanowire-superconductor hybrids \cite{Alicea-2010,Lutchyn-2010} are currently leading candidates for the detection and manipulation of Majorana zero modes (MZMs) \cite{kitaev:physusp2001}, with the ultimate goal of realizing topological quantum computing \cite{Sarma2015,Aasen-2016,obrien:prl2018}. The past few years have witnessed tremendous experimental advancements \cite{Mourik-2012,Deng-2016,Scaling_ZBP_Marcus,NNat_2018,zhang2021large} aimed at the ascertained detection of MZMs via conductance measurements, with the zero bias conductance quantization as a possible tell-tale signature. The zero bias conductance peak (ZBCP) of measure $2e^2/h$ arising out of a coherent and perfect Andreev reflection \cite{Prada-2020} is regarded as a strong signature of the MZM. This signature is expected from a local conductance measurement in a normal - topological superconductor (N-TS) link \cite{Hao_next,Prada-2020,Zhang-2018,zhang2021large}. \\ 
\indent In recent times, there has also been a keen interest in pursuing transport measurements across a three terminal normal-topological superconductor-normal (N-TS-N) configuration \cite{puglia2020closing}, where specific non-local conductance signatures \cite{Akhmerov} could augment the certainty of MZM detection. Before graduating toward the actual manipulation or harnessing of MZMs \cite{Hao_next,Prada-2020}, many unresolved aspects exist, that include the possibility of a non-topological origin of the zero bias conductance as well as understanding the role of interactions within the topological superconductor channel \cite{Stanescu,Andreev_Majorana,Sci_Post_Wimmer,PhysRevB.97.214502,PhysRevB.97.165302,PhysRevB.98.085125,PhysRevB.98.155314,PhysRevB.99.035312,Pan-2020}. \\
\indent  All this has led to a side-by-side strong theoretical push \cite{Stanescu,Andreev_Majorana,Sci_Post_Wimmer,PhysRevB.97.214502,PhysRevB.97.165302,PhysRevB.98.085125,PhysRevB.98.155314,PhysRevB.99.035312,Pan-2020}, where a lot of focus has been on resolving the issue about the origins of the observed ZBCP. In parallel, there is also a strong need to push for theoretical works that go beyond minimal 1D models in order to include multi-mode effects \cite{Prada-2020,PhysRevB.98.125417,Praveen}, realistic potentials \cite{PhysRevX.8.031040,PhysRevX.8.031041,Zero-energy-yeyati,Vulik_2016,Pan-2019-prb,Unified-roman}, the inclusion of disorder \cite{PhysRevB.98.085125,PhysRevB.98.155314,PhysRevB.99.035312,Pan-2020} and also scattering and relaxation effects \cite{DasSarma_Dissipation}.\\ 
\indent  While the importance of ascertaining the topological origin of the observed experimental conductance signatures needs to be emphasized, in this paper, we take the route of analyzing the nature of a topological MZM in a typical three terminal N-TS-N configuration and the corresponding ZBCP, first in the coherent transport regime, and then in the non-coherent regime where dephasing effects due to relaxation mechanisms are present. \\
\indent Theoretical works have been based on minimal 1D models used to analyze the N-TS links \cite{Stanescu,Andreev_Majorana,Sci_Post_Wimmer,PhysRevB.97.214502,PhysRevB.97.165302,PhysRevB.98.085125,PhysRevB.98.155314,PhysRevB.99.035312,Pan-2020} and have typically been based on a scattering formulation of quantum transport \cite{Groth_2014} in the coherent regime. In this work, we use the Keldysh non-equilibrium Green's function (NEGF) technique \cite{Cayao,Cayao_3,Cayao_2,Stoof-Kitaev,Low_Energy_Yeyati,Praveen,Cole,leumer2020linear} to analyze the transport across the N-TS-N junction, with the inclusion of channel interactions in the non-coherent regime \cite{Datta,PhysRevB.75.081301,7571106,PhysRevApplied.8.064014,doi:10.1063/1.5044254}. A three terminal geometry depicted schematically in Fig.~\ref{fig:schematic-a}, comprises a topological superconducting region flanked by two normal contacts, a top gate and tunnel barriers. Such a geometry permits local and non-local transport measurements, with the local Andreev reflection, the crossed Andreev and direct transmissions respectively \cite{Akhmerov,leumer2020linear,Cole}. In the coherent regime, we point out contributions of the local Andreev reflection and the non-local transmissions toward the conductance signatures and that a half of the conductance quantum $(e^2/h)$ is indeed maintained. The half fraction, as pointed out in earlier works \cite{leumer2020linear} is indeed due to a voltage divider formed at the two N-TS junctions, under a symmetric bias situation. The effect of contact broadening clearly shows the maintenance of the ZBCP at zero temperatures \cite{Scaling_ZBP_Marcus,Low_Energy_Yeyati}. In addition, we clarify on the roles of the contact broadenings on the local and the non-local components of the transmission around the parity crossings, and the conductance lineshapes of the MZM. Interestingly, we note that at larger $B$-field parity crossings, the contribution of the Andreev reflection process decreases which is compensated by the non-local processes in order to maintain the conductance quantum regardless of contact coupling strength.\\
\indent We then analyze the non-coherent transport regime by including dephasing due to fluctuating impurities and the resulting momentum randomization processes which allows one to transition smoothly from the coherent ballistic limit all the way to the diffusive limit \cite{camsari2020nonequilibrium}. It is pointed out that the Majorana character of the zero mode is unchanged, as seen in the spatial variation of the wavefunction, and the zero-bias conductance peak is retained. However, we observe a reduction in the peak magnitude that scales with the strength of the impurity potential fluctuations. Important distinctions between dephasing processes in the non-coherent regime and the contact-induced tunnel broadenings on the conductance lineshapes in the coherent regime \cite{Scaling_ZBP_Marcus} are clearly elucidated. Most importantly our results reveal that the addition of dephasing in the set up does not lead to any notable length dependence to the conductance of the zero modes, contrary to what one would expect in a gradual transition to the diffusive limit. We believe that this work paves the way for systematic introduction of scattering processes into the realistic modeling of Majorana nanowire hybrid devices. \\
\indent This paper is organized as follows. In Sec.~\ref{sec:formalism}, we introduce the 1-D Rashba nanowire channel and the NEGF formalism in both the coherent and non-coherent regime. Following this, we describe the main results of coherent transport in Sec.~\ref{sec:coherent}, and that of non-coherent transport in Sec.~\ref{sec:non-coherent}. We also depict the results of the longer wire in Sec.~\ref{sec:long_wire}. Many of the results presented deserve a longer discussion that is dealt with in Sec.~\ref{sec:discussions} and we conclude in Sec.~\ref{sec:conclusions}.
\section{\label{sec:formalism}Set up and formalism}
The transport set up as depicted in Fig.~\ref{fig:schematic-a} comprises the topological superconducting channel, connected to two normal metallic contacts. The schematic also consists of two tunnel gates labeled $V_{G1}$ and $V_{G2}$ for modulating the tunnel barriers and a central gate $V_G$. We consider a one-dimensional semiconducting nanowire with Rashba spin-orbit coupling $\alpha_R$, which is placed in an external axial-Zeeman field $B$ coupled to an s-wave superconductor in order to induce a proximity effect \cite{Hao_next}. Within the Keldysh NEGF formalism, we can translate the set up as depicted in Fig.~\ref{fig:schematic-b}, where the channel is represented via the Hamiltonian and the connection to contacts as well as dephasing via self energies. We begin by detailing the computational approach we employ to pursue the calculations in this paper. 
\subsection{The Rashba nanowire Hamiltonian}
For our computational purposes, we consider the 1-D Rashba Hamiltonian \cite{cayao2017hybrid} in the tight-binding representation in a discrete lattice \cite{Cayao,cayao2017hybrid} with $N$ sites and effective lattice constant $a$, written as 
\begin{equation}\label{eq:H-TB}
    \mathcal{\hat{H}_0}=\sum_{i} c_{i}^{\dagger} \alpha_S c_{i}+\sum_{<i,i+1>} c_{i}^{\dagger} \beta c_{i+1}+ \sum_{<i,i+1>} c_{i+1}^{\dagger} \beta^{\dagger} c_{i},
\end{equation}
where $ c_i= {\begin{bmatrix} d_{i,\uparrow} & d_{i \downarrow} \end{bmatrix}}^{T}$ represents the annihilation operator of the spinor, with individual spin components $d_{i \sigma}$ of spin $\sigma= \uparrow, \downarrow$ on a site $i$. We then have $\mathcal{\hat{H}_0}=\sum_{i,j} c_i^{\dagger} H_0 c_j$, where $H_0$ is the first quantized Rashba nanowire Hamiltonian such that
\begin{eqnarray} 
\alpha_S &=& \left(\begin{array}{cc}2 t-\mu & V_Z \\ V_Z & 2 t-\mu\end{array}\right) \nonumber\\
\beta &=& \left(\begin{array}{cc}-t_0& t_{S O} \\ -t_{S O} & -t_0\end{array}\right) ,
\end{eqnarray}
where Zeeman energy $V_Z =g \mu_{B} B/2$, where $g$ is the electronic g-factor, $\mu_B$ is the Bohr magneton, and $B$ is the applied axial magnetic field. The tight-binding hopping parameter $t_0=\frac{\hbar^2}{2m^{*}a^2}$ with $m^{*}$ being the effective mass and $\hbar$ being the reduced Planck's constant, and $t_{SO}=\frac{\alpha_R}{2a}$, where $\alpha_R$ is the Rashba spin-orbit coupling term. $\mu$ is the chemical potential in the nanowire, which is taken to be spatially uniform. Here $\alpha_S$ represents the on-site $2 \times 2$ matrix and $\beta$ represents the $2 \times 2 $ hopping matrix in the discretized tight binding representation. \\
\indent In the Bogoliubov-de Gennes (BdG) representation for the superconducting state, we can write the Hamiltonian of the Rashba nanowire in its 4-component site Nambu spinor $\Psi_i= {\begin{bmatrix} d_{i,\uparrow} & d_{i \downarrow} & d_{i \uparrow}^{\dagger} & d_{i \downarrow}^{\dagger} \end{bmatrix}}^{T}$ representation as $\mathcal{H}= \frac{1}{2}\sum_{i,j} \Psi^{\dagger}_i H_{BdG} \Psi_j$, where
\begin{equation} \label{eq:H-BdG}
    H_{BdG}=\left(\begin{array}{cc}H_{0} & \Delta_{m} \\
    \Delta^{\dagger}_{m} & -H_{0}^{*} \end{array}\right)
\end{equation}
with the pairing potential given as $\Delta_{m} = i \Delta ( \mathbbm{1}_N \otimes \sigma_y)$, where $\Delta$ is the superconducting order parameter induced in the nanowire by proximity effect and $\sigma_y$ is the $y$ Pauli matrix. Increasing the magnetic field $B$ (and thereby $V_Z$) in the wire drives the system into the topological phase beyond the critical Zeeman energy $V_Z^c = \sqrt{\Delta^2 + \mu^2}$, which corresponds to a critical magnetic field $B_c$.\cite{Alicea-2010, Lutchyn-2010}\\
\indent We take the contacts to be in the eigenbasis with their Hamiltonians given by
\begin{equation}
\hat{H_C}=\sum_{k \sigma, \alpha \in L,R} \epsilon_{\alpha k \sigma} c_{\alpha k \sigma }^{\dagger} c_{\alpha k \sigma },
\label{eq:HamC}
\end{equation}
where $c_{\alpha k \sigma}^{(\dagger)}$ represents the annihilation (creation) operator of an electronic momentum eigenstate of energy $\epsilon_{\alpha k \sigma}$ indexed with $k$ of spin $\sigma =\uparrow, \downarrow$ in contact labeled $\alpha = L,R$. Since the contacts are normal metal, the energy levels are spin degenerate such that $\epsilon_{\alpha k \uparrow}=\epsilon_{\alpha k \downarrow}= \epsilon_{\alpha k}$ \\
\indent The coupling between the contacts and the topological superconductor is represented by the spin-conserving tunneling Hamiltonian given by
\begin{equation}
\hat{H_T}=\sum_{k \sigma m \alpha \in L,R}  \left [ \tau_{\alpha k m \sigma} c_{\alpha k \sigma}^{\dagger} d_{m \sigma} + {\tau}^{*}_{ \alpha k \sigma m} d_{m \sigma}^{\dagger} c_{ \alpha k \sigma} \right ],
\end{equation}
where $\tau_{\alpha k m \sigma}$ is an element of the tunnel coupling matrix between a state $k$ in contact $\alpha=L,R$ with spin $\sigma = \uparrow, \downarrow$ and a site $m$ with the same spin $\sigma = \uparrow, \downarrow$ in the channel. In this paper, and of relevance to the Majorana nanowire, we take a spin-independent tunnel coupling $\tau_{\alpha k m \uparrow}= \tau_{\alpha k m \downarrow}= \tau_{\alpha k m}$.  We now proceed to describe the transport formulation that involves the proximitized Rashba nanowire system as the central system coupled to two normal contact as leads $\alpha =L,R$. Furthermore, referring to Fig.~\ref{fig:schematic-b}, we note that $m=1 (m=N)$ when $\alpha=L (\alpha=R)$, implying that $\tau_{Lk} = \tau_{Lk1}$ and $\tau_{Rk} = \tau_{RkN}$.
\subsection{\label{sec:negf}The Keldysh NEGF formalism}
Though the foundations of scattering theory \cite{Groth_2014} allow us to evaluate various quantities and currents in various structures, a systematic computational framework for arbitrary geometries including unconventional pairing \cite{PhysRevB.57.10972} may be set up using the Keldysh non-equilibrium Green's function (NEGF) technique \cite{Jauho,Datta}. The formalism also allows one to capture intra-channel interactions in a systematic way, in principle facilitating a smooth transition from a purely quantum ballistic regime all the way toward the diffusive regime \cite{Datta,PhysRevB.75.081301,camsari2020nonequilibrium}. This aspect of the inclusion of dephasing will also be dealt with in this paper. First, we will cover the coherent ballistic regime, in which the Landauer-B\"uttiker type expressions for the conductance signatures of the MZMs will be derived. 
Based on the Hamiltonian from \eqref{eq:H-BdG}, for steady state transport calculations, we define the retarded Green's function in the matrix representation as
\begin{equation}
[G^{r}(E)] = \left [ (E+i \eta)I - H_{BdG} - \Sigma^{r}_L - \Sigma^{r}_R - \Sigma^{r}_s \right ]^{-1},
\label{Main_Ret}
\end{equation}
where $E$ is the energy free variable and $I$ is the identity matrix of the dimension of the Hamiltonian. The quantities $\Sigma^{r}_{\alpha}, \Sigma_s^r$ represent the retarded self energies of the contacts $\alpha$, and the scattering processes respectively. The latter will be considered in the non-coherent regime. To evaluate transport properties, one also requires the {\it{lesser}} Green's function, which is given by a matrix product 
\begin{equation}
[G^{<}(E)] = [G^r] [\left ( \Sigma_L^{<} + \Sigma_R^{<} + \Sigma_s^{<} \right )] [G^a], 
\label{Sigma_Less}
\end{equation}
where $\Sigma_{\alpha}^{<}$ is the lesser self energy of the contacts are defined in Appendix in \eqref{eq:Sigma_less}.
 \\
\indent The calculation of retarded contact self energies $[\Sigma^{r}_{L(R)}]$ can be done in any basis of the contacts depicted in Fig.~\ref{fig:schematic-b}. $\Gamma_{L(R)} = i \left([\Sigma^{r}_{L(R)}] - [\Sigma^{r}_{L(R)}]^{\dagger}\right)$ denotes the broadening matrix of either lead.  Here, we consider the contacts in the eigenbasis of the Hamiltonian in \eqref{eq:HamC}, where the self energy matrix is diagonal. In the wide band approximation described in Appendix \ref{sec:NEGF-appendix}, the contact broadening matrix is real and independent of energy, so that the self energy matrix can be written as $[\Sigma^{r}_{L(R)}]=-i [\Gamma_{L(R)}]/2$. Further,  $\Sigma^{r}_{L}(i,i) = -i \gamma_L/2$ for $i=1$ to $i=4$ and $\Sigma^{r}_{R}(i,i)= -i \gamma_R/2$ for $i=4N-3$ to $i=4N$, and all other diagonal elements are zero.  These represent the up and down-spin electron (hole) submatrices at each  site $i$ in the 4-component site Nambu representation. The energy-independent self energy is given by $\gamma_{L(R)}= 2\pi \int d \epsilon_{k \alpha} \mid \tau_{\alpha,k} \mid^2 \delta(E-\epsilon_{k \alpha})$, which we treat as a parameter here. In this paper we take $\gamma_L=\gamma_R = \gamma$. \\
\indent With the evaluation of the retarded Green's function, one can calculate the spectral function and the density of states as follows:
\begin{equation}
[A(E)] = i[(G^r-G^a)] = [G^r][\Gamma][G^a], 
\label{eq:spec_main}
\end{equation}
where, the diagonal elements of the spectral function $D(E) = A(E)/2 \pi$ are related to the local density of states (LDOS), and $[G^a(E)] = \left [ G^r(E) \right ]^{\dagger}$ represents the advanced Green's function. \\
{\it{Coherent transport:}}  The current operator \cite{Datta} from the Keldysh NEGF formalism can be derived from the {\it{lesser}} Green's function in Nambu space, as detailed in Appendix A as follows:
\begin{widetext}
\begin{equation}
I^{op}_L(E)= \frac{1}{2}\left[\frac{e}{h} Tr \left(\tau_z \left [ [G^r] [\Sigma_L^{<}] - [\Sigma_L^{<}] [G^{a}]+  [G^<] [\Sigma_L^{a}] -[\Sigma_L^{r}][]G^<] \right ] \right ) \right ],
\label{Full_Iop_Nambu}
\end{equation}
\end{widetext}
where $\tau_z= I \otimes \sigma_z$, with $\sigma_z$ being the Pauli-$z$ matrix, representing the fact that the net current is the difference between the currents of the composite electron and hole blocks. The factor of half comes from the BdG representation of the original Hamiltonian. Following the detailed derivation in Appendix A, we can obtain the electron (hole) current across the device as a sum of three components:
\begin{widetext}
\begin{eqnarray}
I^{e(h)}_L(E) = \frac{e}{h} \left (Tr\left ( \Gamma_L^{ee (hh)}G^r \Gamma_R^{ee (hh)} G^a \right ) \left [ f_L^{ee (hh)}(E)-f_R^{ee (hh)}(E) \right ] \right ) \label{direct_main} \\
\qquad +  \frac{e}{h}\left (Tr\left ( \Gamma_L^{ee (hh)}G^r \Gamma_L^{hh (ee)} G^a \right ) \left [ f_L^{ee (hh)}(E)-f_L^{hh (ee)}(E) \right ] \right ) \label{Andreev_main} \\
\qquad + \frac{e}{h} \left ( Tr\left ( \Gamma_L^{ee (hh)}G^r \Gamma_R^{hh (ee)} G^a \right ) \left [ f_L^{ee (hh)}(E)-f_R^{hh (ee)}(E) \right ] \right ),  \label{Crossed_Andreev_main}
\end{eqnarray}
\end{widetext}
where the superscripts $ee ~(hh)$ represent the block diagonal sector of the corresponding matrices in the electron-hole Nambu space, and \eqref{direct_main} represents the direct transmission process of either the electron or the hole, \eqref{Andreev_main} represents the direct Andreev transmission and \eqref{Crossed_Andreev_main} represents the crossed Andreev transmission. The three components $T_D(E)$, $T_A(E)$ and $T_{CA}(E)$ corresponding to direct transmission, Andreev reflection and and crossed Andreev transmission processes shown above are
\begin{eqnarray}
    T_D(E) = Tr\left ( \Gamma_L^{ee (hh)}G^r \Gamma_R^{ee (hh)} G^a \right ) \\
    T_A(E) = Tr\left ( \Gamma_L^{ee (hh)}G^r \Gamma_L^{hh (ee)} G^a \right ) \\
    T_{CA}(E) = Tr\left ( \Gamma_L^{ee (hh)}G^r \Gamma_R^{hh (ee)} G^a \right ).
\end{eqnarray}
At this point it is worth noting that $f_{\alpha}^{ee}(E)  = f(E-\mu_{\alpha}), f_{\alpha}^{hh}(E)=f(E+\mu_{\alpha})$, representing the Fermi-Dirac distributions of the electronic and hole sector respectively. In these calculations, we must recall that the up and down-spin degrees of freedom in the electron and hole sectors respectively are included in \eqref{direct_main}, \eqref{Andreev_main} and \eqref{Crossed_Andreev_main}. 
\begin{figure}[!htb]
\includegraphics[width=3in,height=3in]{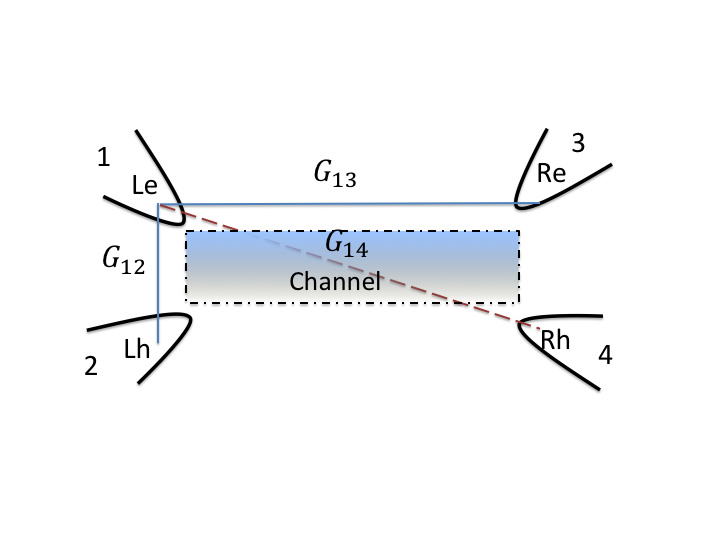}
\caption{Equivalent Landuer-B\"uttiker network for our device. Each contact is further divided into ``two contacts'', the electron part and the hole part, namely $Le$, $Lh$, $Re$ and $Rh$ respectively.}
\label{Figure_LB}
\end{figure}
\begin{figure}%
	\subfigure[]{ \includegraphics[height=0.35\textwidth,width=0.45\textwidth]{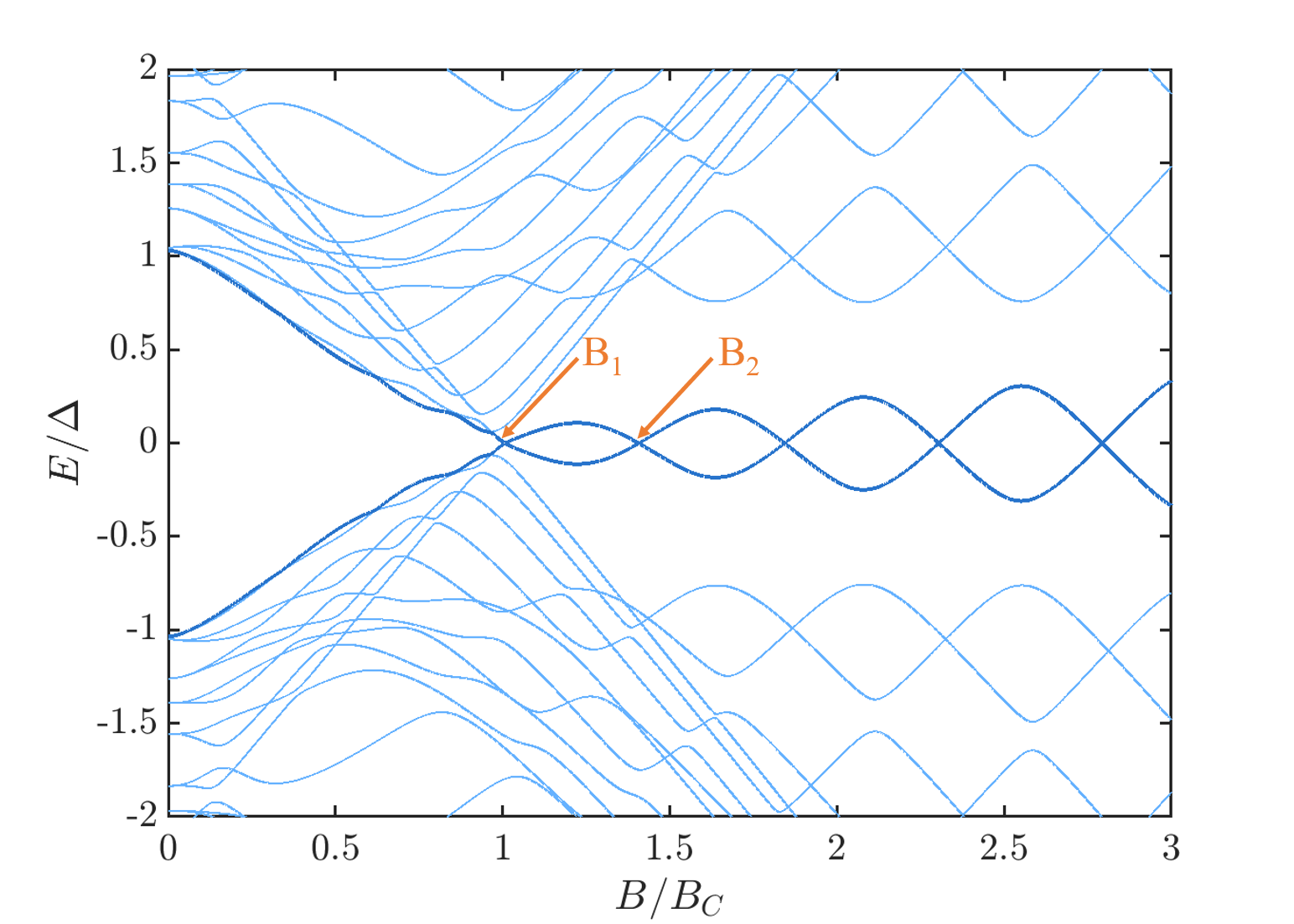}\label{fig:spectrum-short}}
	\quad
	\subfigure[]{ \includegraphics[height=0.35\textwidth,width=0.45\textwidth]{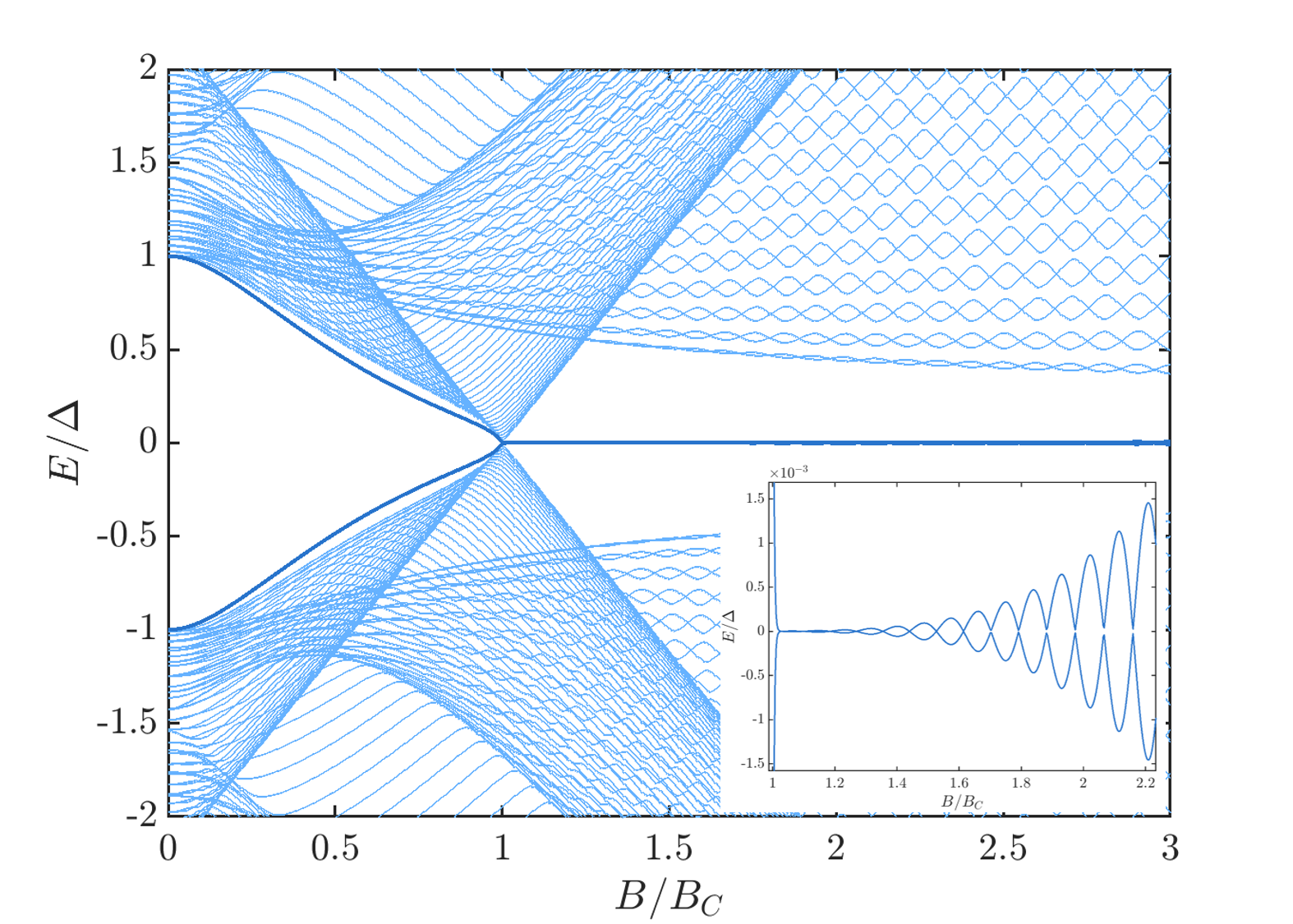}\label{fig:spectrum-long}}
	\quad
	\caption{\label{fig:spectra} Energy spectrum of a short (a), $l=1.6 \mu m$) and a long (b), $l=8 \mu m$) Majorana nanowire with varying normalized Zeeman field $B/B_c$. The gap closing and re-opening at  $B=B_c$ is seen clearly confirming the topological regime for $B \geq B_c$. The oscillations in energy are due to hybridization of the zero-energy Majorana modes at either ends in a finite nanowire. While these oscillations are seen prominently for the short wire, their amplitude is suppressed for longer wires where the Majorana localization length $l_M \ll l$ (but are still present, as seen in the inset). The points marked $B_1$ and $B_2$ in (a) represent the first two distinct zero energy parity crossings for the shorter wire. }
	\label{fig:spectrum}
\end{figure}
\indent A somewhat rudimentary understanding of the above results can also be sought from the Landauer-B\"uttiker formalism. Even though our device has two contacts, we conceptually have a four terminal device as sketched in Fig.~\ref{Figure_LB}. The current across such a multi-terminal device \cite{Datta} as seen from a terminal $p$ can be written as 
\begin{equation}
I_{p}(E) = \frac{1}{e}\left [\sum_{q} G_{qp}(E) f_{p}(E)- G_{pq}(E)f_{q} (E)\right ],
\label{eq:LB}
\end{equation}
where $G_{pq}$ represents the conductance of the segment $q \rightarrow  p$, and $f_{p(q)}$ refers to the Fermi-Dirac distribution in terminal $p(q)$, with chemical potential $\mu_{p(q)}$. In the 1D channel that we consider, the conductances obey Onsager symmetry such that $G_{qp}=G_{pq}$, and hence $I_{p} = \sum_{q} G_{pq}\left ( f_{p}-f_{q} \right )$. Each contact has its electronic and hole component labeled as $Le (1)$, $Lh (2) $, $Re (3)$ and $Rh(4)$ respectively. Now we can write the terminal current for (say) the $Le (1)$ contact as 
\begin{equation}
I_1 = G_{12}(f_1-f_2) + G_{13} (f_1- f_3) + G_{14}(f_1-f_4),
\label{eq:LB_1}
\end{equation}
where the energy argument is understood as described in \eqref{eq:LB}. 
Now comparing the above equation with \eqref{direct_main}, \eqref{Andreev_main} and \eqref{Crossed_Andreev_main}, we note at zero temperature that $G_{12} = \frac{e^2}{h} T_A(E=0)$, $G_{13} = \frac{e^2}{h} T_D(E=0)$ and $G_{14} = \frac{e^2}{h} T_{CA}(E=0)$, representing the Andreev transmission representing the local conductance, the direct and the crossed Andreev transmissions that comprise non-local conductance respectively. Throughout this work, we work in the zero temperature limit with the thermal energy much less than the other energy scales involved in the system ($k_B T \ll \Delta$). \\
\indent For reasons involving current conservation in systems featuring superconducting regions, well noted in previous works \cite{Yeyati,Lim_2012,Stoof-Kitaev,leumer2020linear}, we resort to a symmetric voltage bias, i.e., $\mu_L= eV/2$ and $\mu_R=-eV/2$.  However, the issue of current conservation can only be via a self consistent calculation of the order parameter as noted in \cite{Yeyati}, or other number conserving methods beyond the mean-field description \cite{Lapa}. We leave such a calculation for future work. \\
{\it{Non-coherent transport:}} In the case of non-coherent transport, intra-channel interactions such as electron-electron interactions, dephasing and inelastic scattering can be included via the scattering self energy $\Sigma^r_s$. In this paper, we include intra-channel interactions in the form of channel fluctuating impurities with localized potentials $U(r_i)$ at the sites, such that their correlator $\bar{D}(i,j) = \left \langle U(r_i) U^{*}(r_i) \right \rangle$ can be used to find the scattering self energies within the self consistent Born approximation \cite{DANIELEWICZ1984239,Datta,PhysRevB.75.081301,doi:10.1063/1.5023159,PhysRevB.98.125417,Praveen}
\begin{eqnarray}
\Sigma_s^{r}(i,j) = \bar{D}(i,j) G^{r}(i,j) \\
\Sigma_s^{<}(i,j) = \bar{D}(i,j) G^{<}(i,j),
\label{scb}
\end{eqnarray}
where, the superscripts $r(<)$ stand for the usual retarded (lesser) Green functions. In this paper, we adopt a homogeneous model with uniform, elastic, and spatially uncorrelated interactions, resulting in a diagonal form of $\bar{D}(i,j) = d_0 \delta_{ij}$. This model discards the off-diagonal elements of the Green’s
function, thus relaxing both the phase and momentum of quasiparticles in the nanowire. \cite{DANIELEWICZ1984239,Datta}. The quantity $d_0$ is the dephasing parameter which represents the magnitude squared of the
fluctuating scattering potentials. The parameter $d_0$ can be modulated so that by gradually increasing it, one can transition from the coherent ballistic limit to the diffusive limit.
This model can also be extended to include non-local fluctuations via the spatial correlations of the impurity potentials \cite{DANIELEWICZ1984239,Datta}. \\
\indent Using \eqref{scb} solved self consistently with the equations for the retarded (lesser) Green's functions given in \eqref{Main_Ret} and \eqref{Sigma_Less} we next obtain the currents given by \eqref{Full_Iop_Nambu}. We must importantly note that here one cannot write the current operator in a compact Landauer-B\"uttiker form and hence cannot separate the local and non-local contributions. 
\begin{table}[]
\caption{\label{tab:table1}%
These parameters \cite{Mourik-2012} are used in all further analysis, unless otherwise stated.}
\begin{ruledtabular}
\begin{tabular}{lll}
Parameter&&
Value\\
\colrule
Effective mass & $m^\ast$ & 0.015 $m_e$ \\
Induced order parameter & $\Delta$ & 0.25 meV \\
Tight-binding hopping parameter &$ t_0$& 10 meV \\
Rashba spin-orbit coupling strength & $\alpha_R$ & 20 meV nm \\
Chemical potential & $\mu$ & 0.5 meV \\
Contact coupling energy & $\gamma$ & 0.025 meV 
\\
Infinitesimal damping parameter & $\eta$ & 10$^{-10}$ eV \\
\end{tabular}
\end{ruledtabular}
\end{table}
\section{\label{sec:results}Results}
\subsection{\label{sec:spectrum}Energy Spectra}
We begin by reiterating some well known results \cite{cayao2017hybrid} in connection with the equilibrium energy spectra of the wires that we consider in this paper. The energy spectrum of the system is evaluated by numerically diagonalizing the discretized tight-binding Hamiltonian \eqref{eq:H-BdG} using the parameter set described in Tab. I . The energy eigenvalues with varying magnetic field $B$ are shown in Fig.~\ref{fig:spectra} for two different lengths of the nanowire: $l\approx 1.6 \mu m$ (Fig.~\ref{fig:spectrum-short})  and $l\approx 8 \mu m$ (Fig.~\ref{fig:spectrum-long}), corresponding to $100$ and $500$ sites respectively in the tight-binding model. The two eigenvalues closest to zero energy are highlighted. The energy gap at decreases as the Zeeman field is increased, finally closing at the topological transition point corresponding to $B=B_c$. Beyond $B=B_c$, topological Majorana bound states are formed at each end of the nanowire. \\
\begin{figure}
    \subfigure[]{
    \includegraphics[height=0.35\textwidth,width=0.45\textwidth]{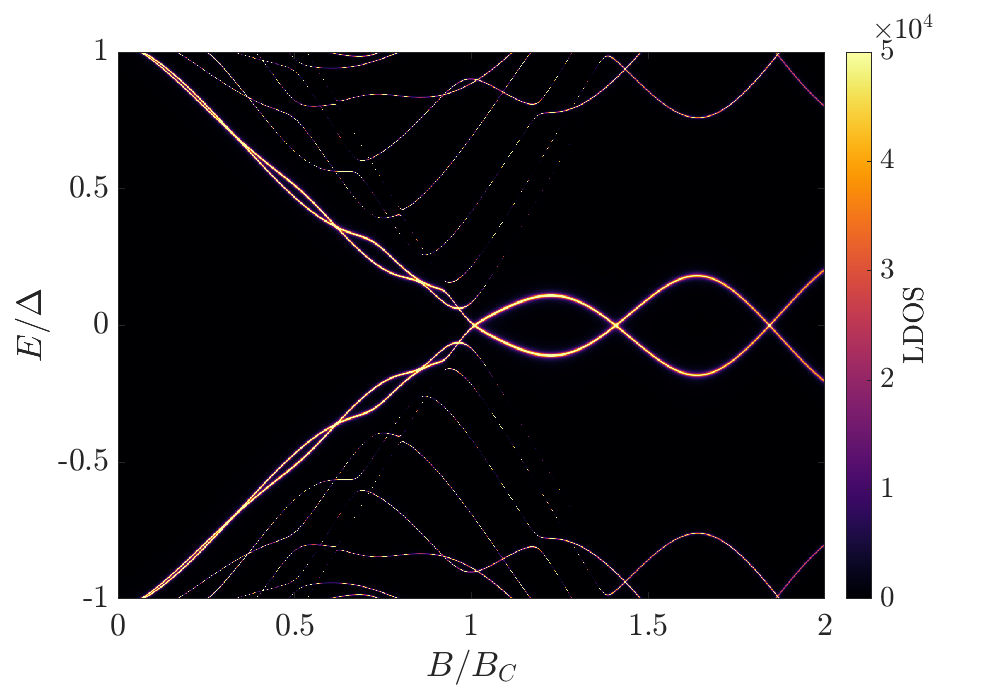}
    \label{fig:coherent-ldos-site5}}
\hfill
\subfigure[]{
    \includegraphics[height=0.35\textwidth,width=0.45\textwidth]{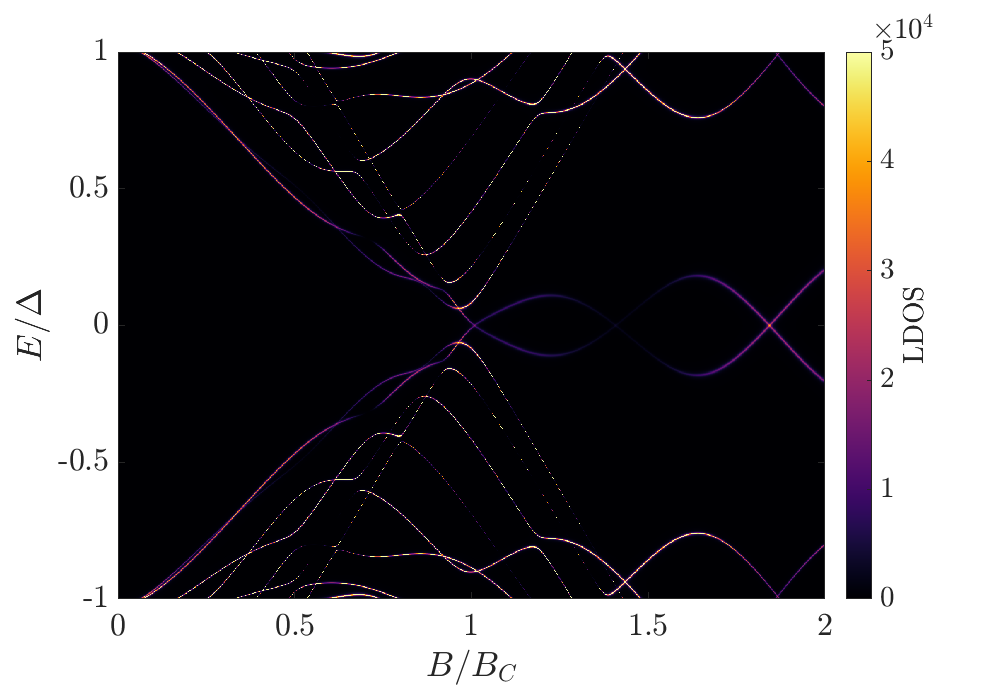}
    \label{fig:coherent-ldos-site50}
    }
\caption{\label{fig:coherent-ldos-site}
Local density of states map with varying magnetic field $B$ at a particular lattice site close to the edge on site 5 (a) and at the center of the nanowire at site 50 (b). We clearly see that the zero-mode contribution to the density of states is mainly from the edge sites as compared to the center, where the zero-mode local density of states contribution is much smaller.
}
\end{figure}
\indent These zero energy states decay exponentially into the bulk of the wire, with their spatial spread characterized by the Majorana localization length $l_M$. The Majorana localization length has a non-trivial dependence on the parameters chosen \cite{cayao2017hybrid,Loss_2012}, and may increase or decrease with the applied magnetic field depending on the spin-orbit length scale. For the range of realistic parameters chosen, $l_M$ indeed increases with increasing Zeeman field, following a minima close to the critical field $B_c$ \cite{cayao2017hybrid}. \\
\indent In a finite nanowire, there is non-zero overlap between the wavefunctions of the the states, and this hybridization leads to oscillations in the corresponding energy levels. The length of the nanowire $l$ relative to the localization length $l_M$ determines the amplitude of these oscillations; however the oscillations are always present, as depicted in the inset of Fig.~\ref{fig:spectrum-long}. For long enough nanowires, the broadening in the energy levels due temperature or coupling to contacts would be larger than the amplitude of these oscillations at Zeeman energies close to the topological transition, and hence the states are effectively zero modes. 
\begin{figure}
    \subfigure[]{
    \includegraphics[height=0.35\textwidth,width=0.45\textwidth]{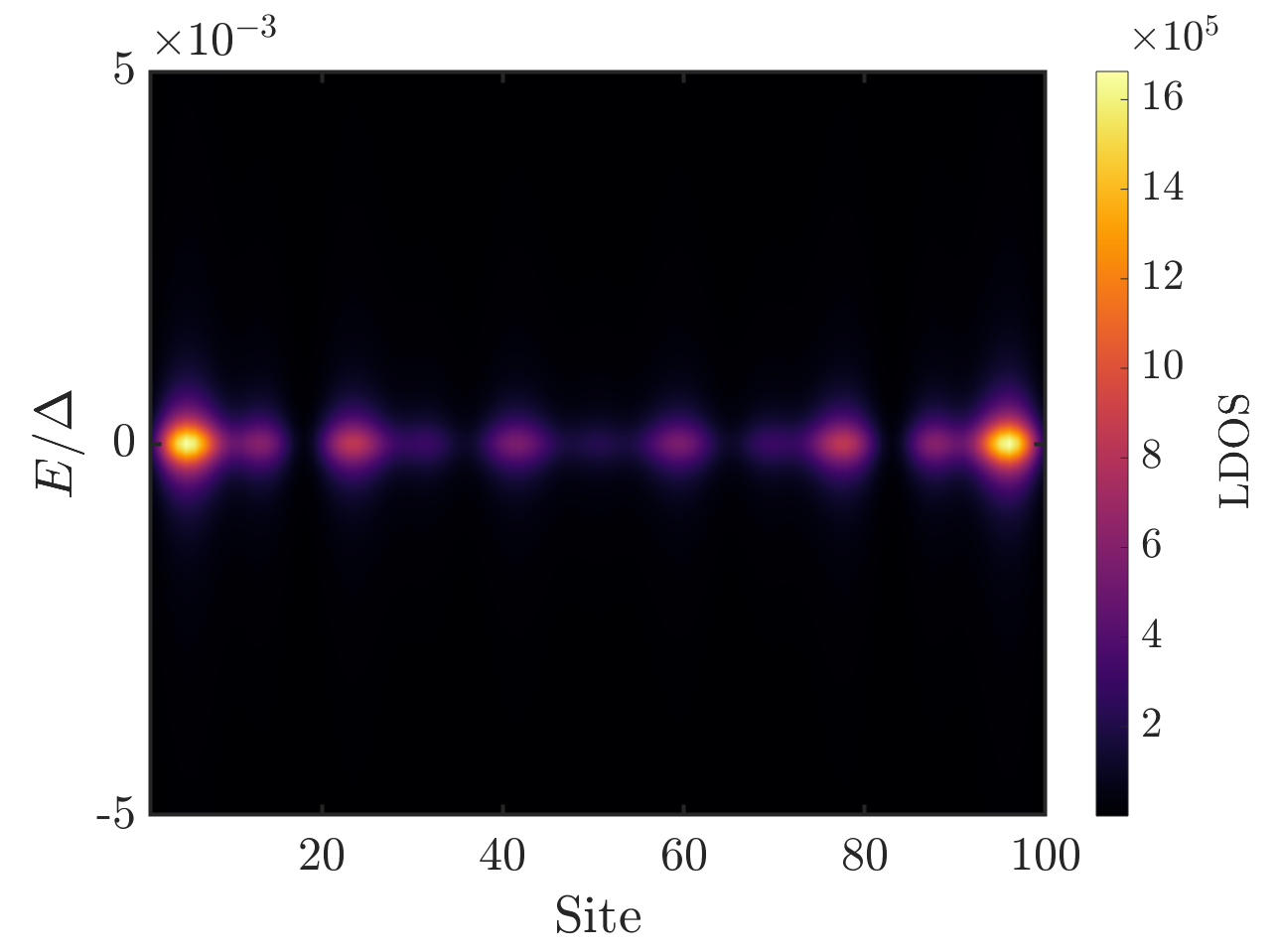}
    \label{fig:coherent-ldos-B1}}
\hfill
\subfigure[]{
    \includegraphics[height=0.35\textwidth,width=0.45\textwidth]{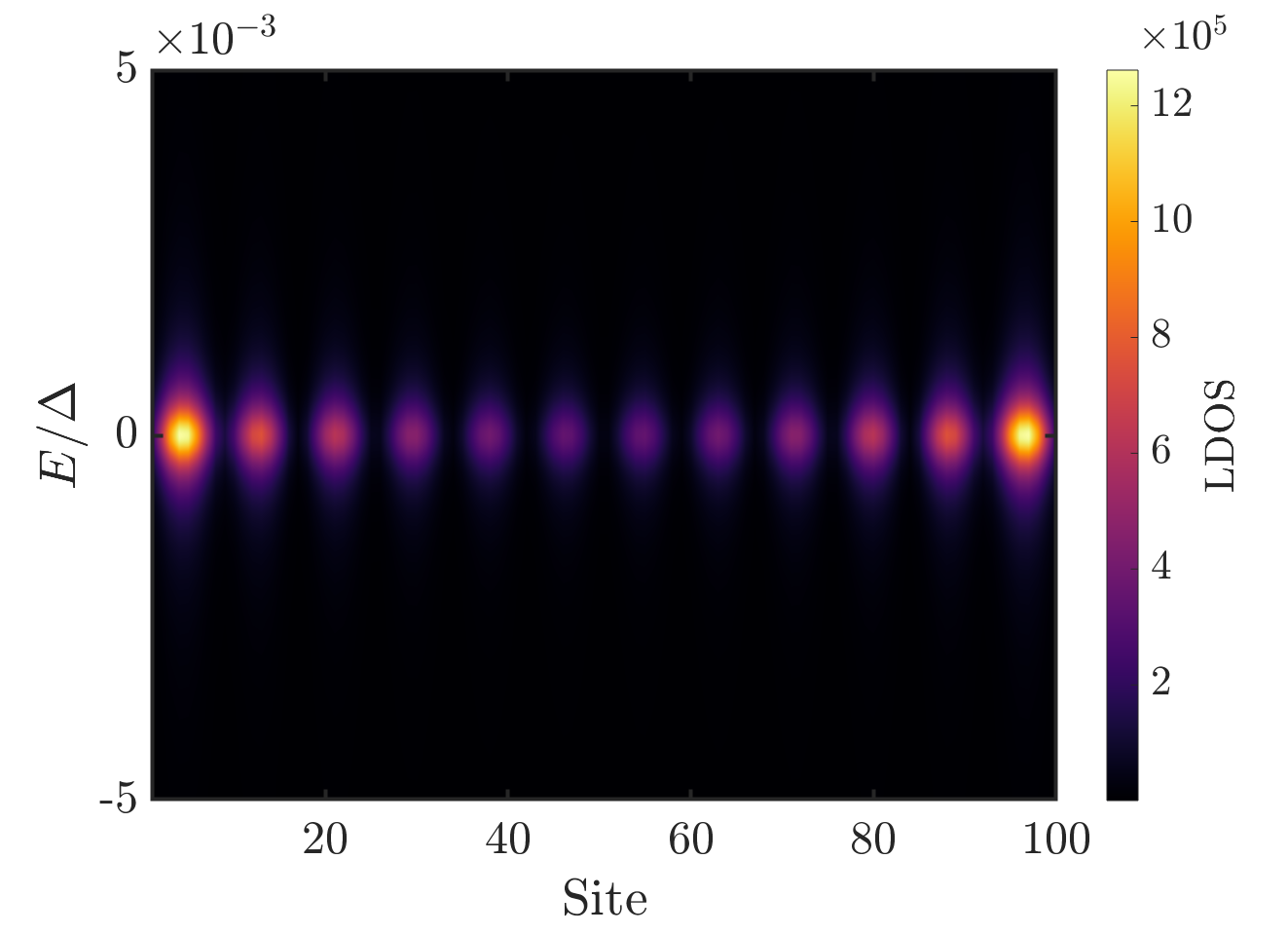}
    \label{fig:coherent-ldos-B2}
    }
\caption{\label{fig:coherent-ldos-B}
Local density of states at the Zeeman energies $B_1$ (Fig.~\ref{fig:coherent-ldos-B1}) and $B_2$ (Fig.~\ref{fig:coherent-ldos-B2}) where the oscillations meet to give exactly zero-energy modes, as marked in Fig.~\ref{fig:spectrum-short}. These modes are clearly seen to be localized at the ends of the nanowire at both values of $B$. The localization is greater at $B_1$ compared to $B_2$ due to a shorter Majorana localization length $l_M$.
}
\end{figure}
\begin{figure*}
\subfigure[]{
    \includegraphics[width=\columnwidth,height=0.76\columnwidth]{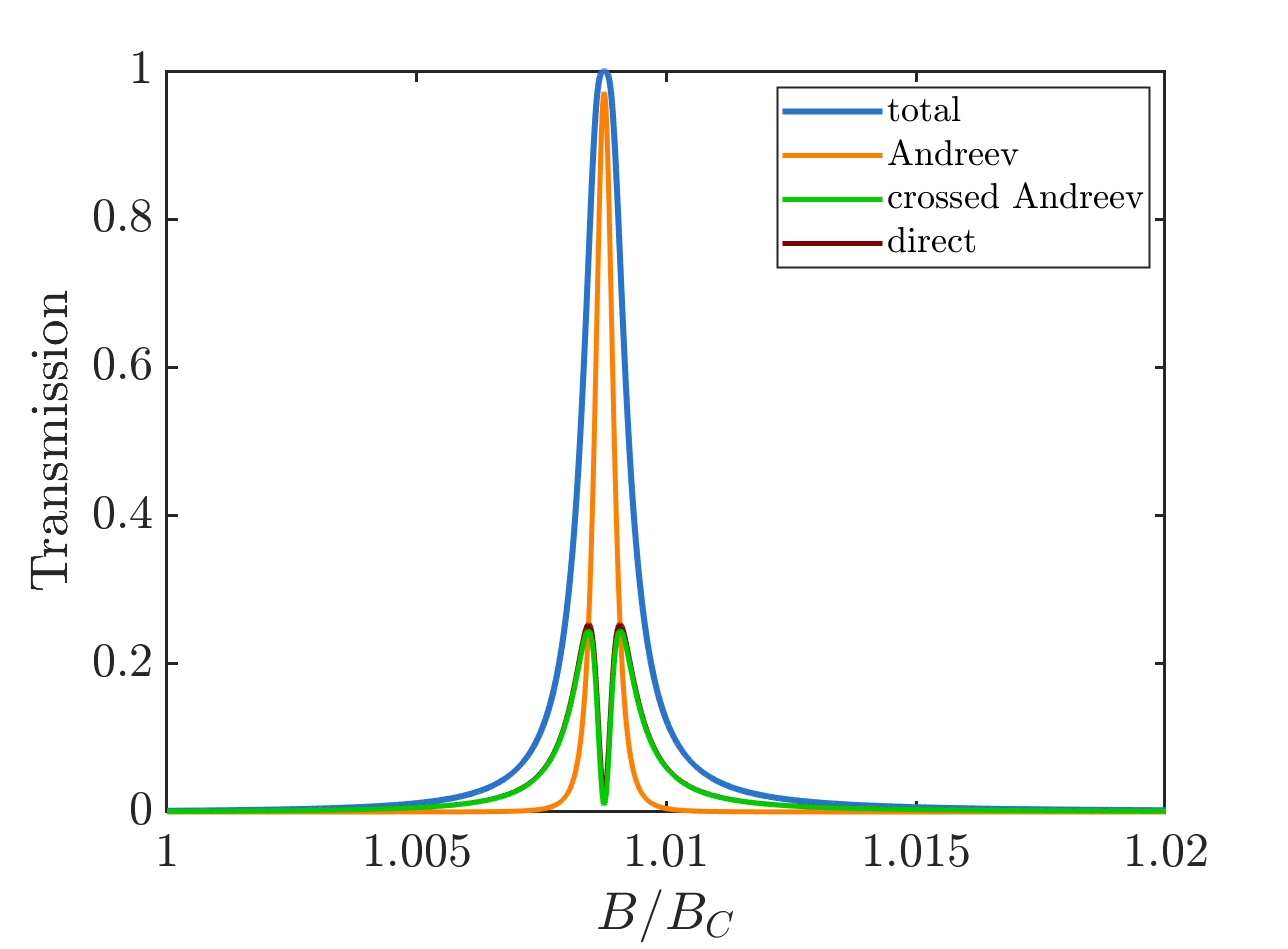}
    \label{fig:coherent-trans-vs-B1}}
\hfill
\subfigure[]{
    \includegraphics[width=\columnwidth,,height=0.76\columnwidth]{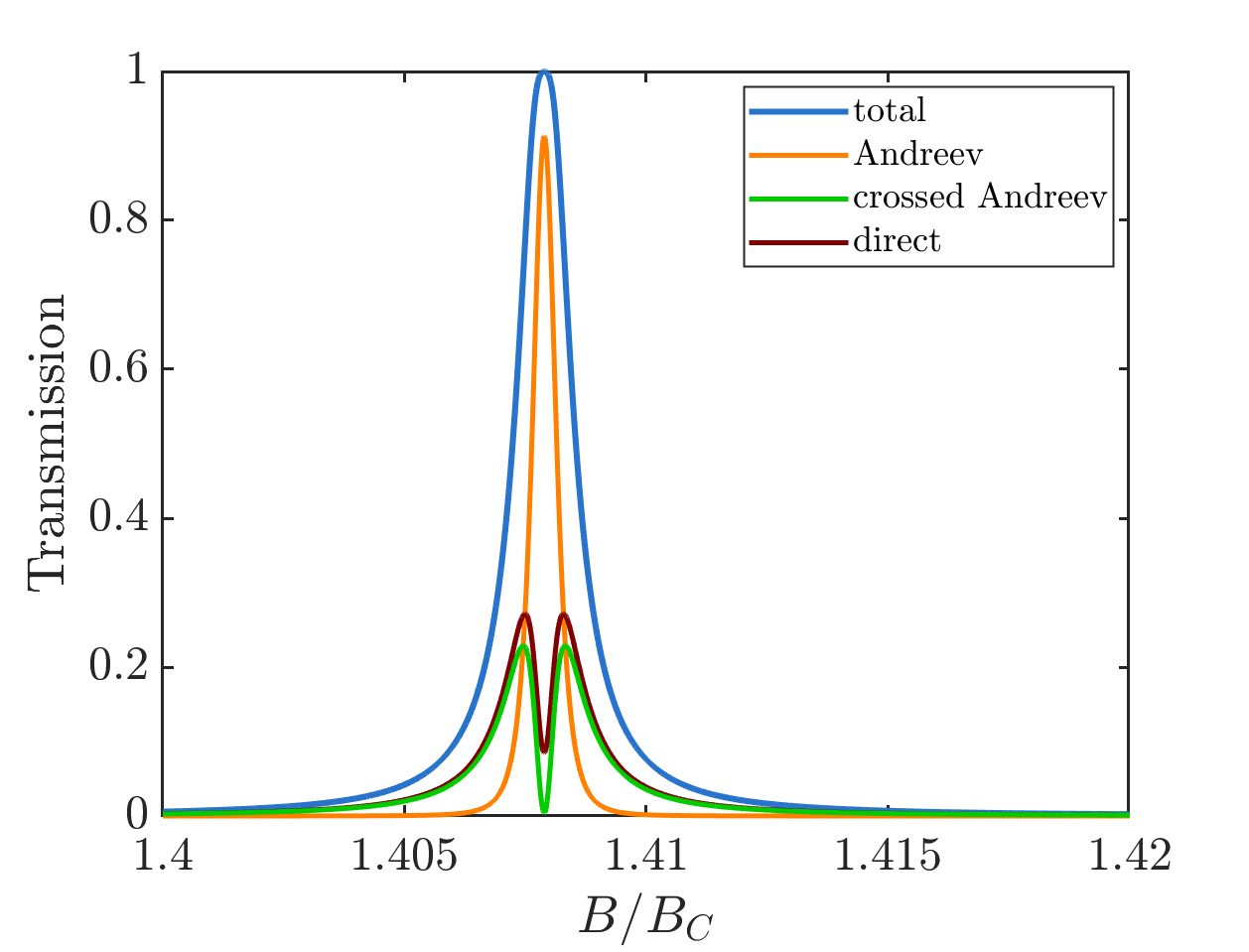}
    \label{fig:coherent-trans-vs-B2}}
\subfigure[]{
    \includegraphics[width=\columnwidth,height=0.76\columnwidth]{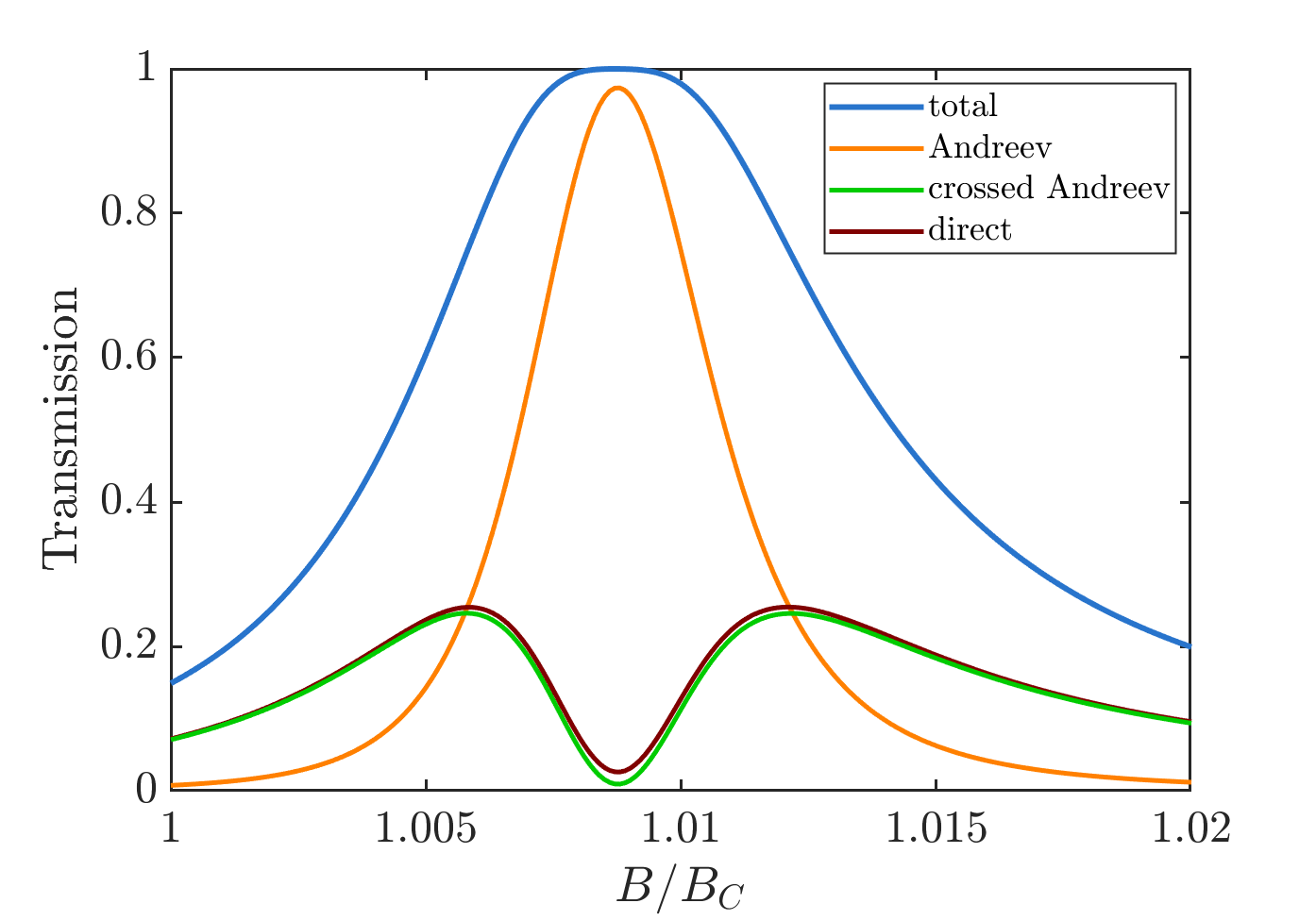}
    \label{fig:coherent-trans-vs-B1-gam2}}
\hfill
\subfigure[]{
    \includegraphics[width=\columnwidth,height=0.76\columnwidth]{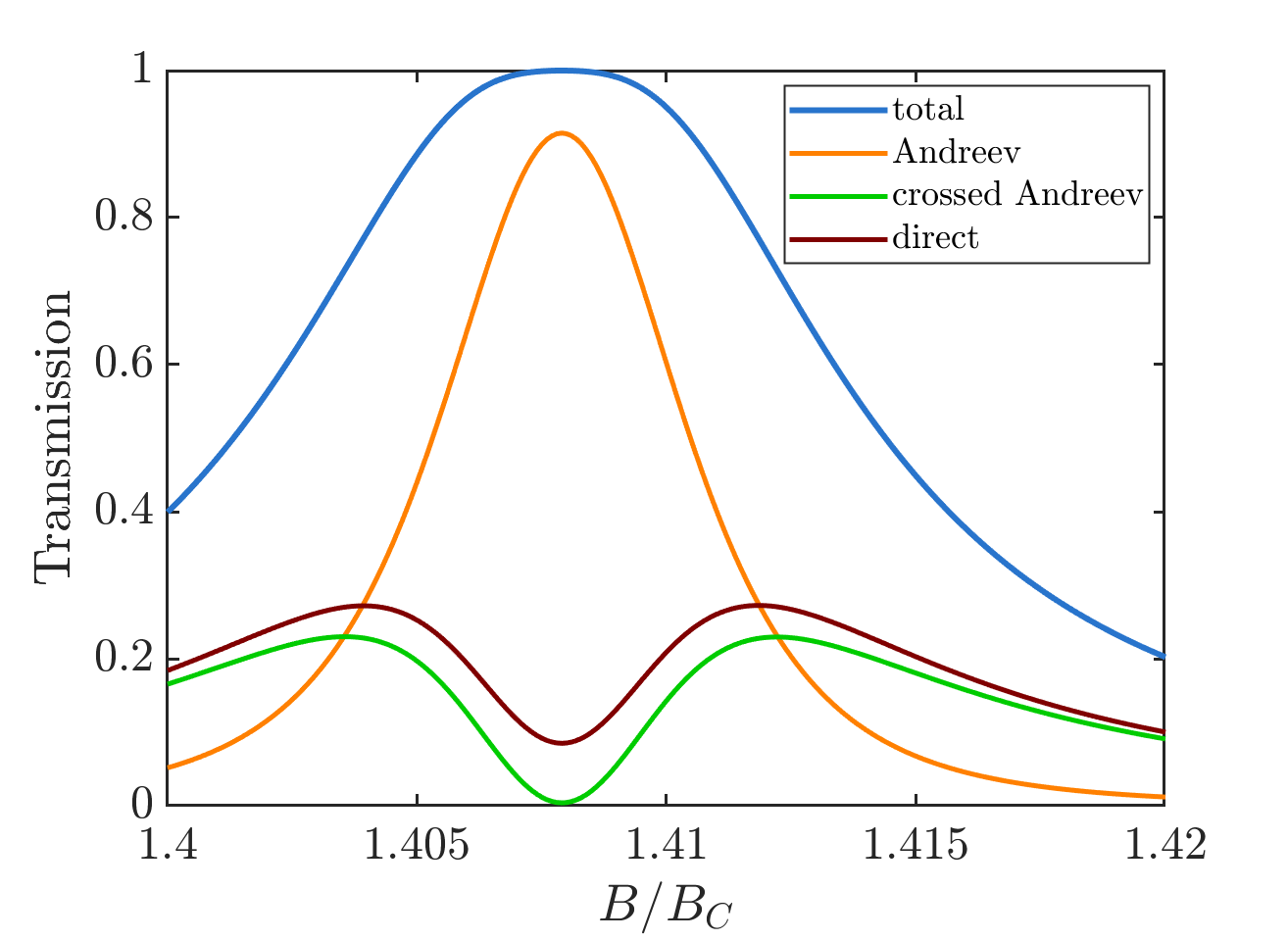}
\label{fig:coherent-trans-vs-B2-gam2}}
\caption{Local and non-local transmission at zero energy with varying Zeeman field. Plots are resolved into the direct, Andreev and crossed Andreev process components, with (a) and (b) evaluated for $\gamma = 0.1 \Delta$ around the points $B_1$ and $B_2$ (Fig. ~\ref{fig:spectrum-short}) respectively, (c) and (d) are evaluated for $\gamma = \Delta$ at the points $B_1$ and $B_2$. In all cases, at the Zeeman energies with exact zero-modes, the direct and crossed Andreev contributions (non-local processes) drop and only the local Andreev process makes the maximum contribution to the transmission peak with a total magnitude of unity. For larger $\gamma$, the peaks are broadened out in energy, and therefore the peak is more robust to small changes in B around $B_1$ or $B_2$, however, retaining the unity maximum. It is also worth noting by comparing (a) and (c) with (b) and (d) that the individual contributions of the local and non-local processes change at the two unique parity crossings, while the total magnitude remains quantized at unity at the magnetic field corresponding to either crossing. }
\label{fig:coherent-trans-vs-B}
\end{figure*}

\begin{figure}
    \subfigure[]{
    \includegraphics[height=0.35\textwidth,width=0.45\textwidth]{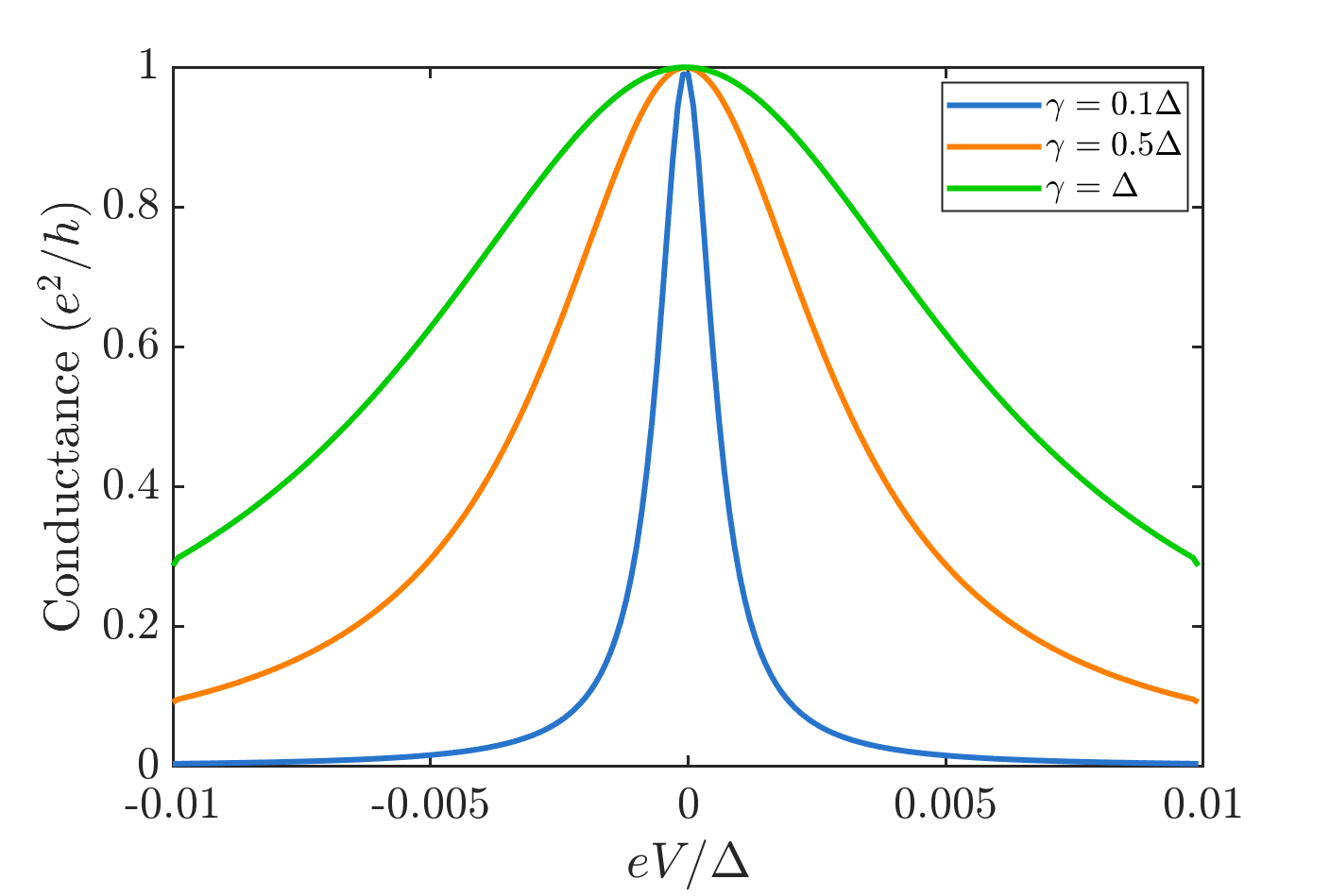}
   \label{fig:coherent-cond-vs-V}}
\hfill
\subfigure[]{
    \includegraphics[height=0.35\textwidth,width=0.45\textwidth]{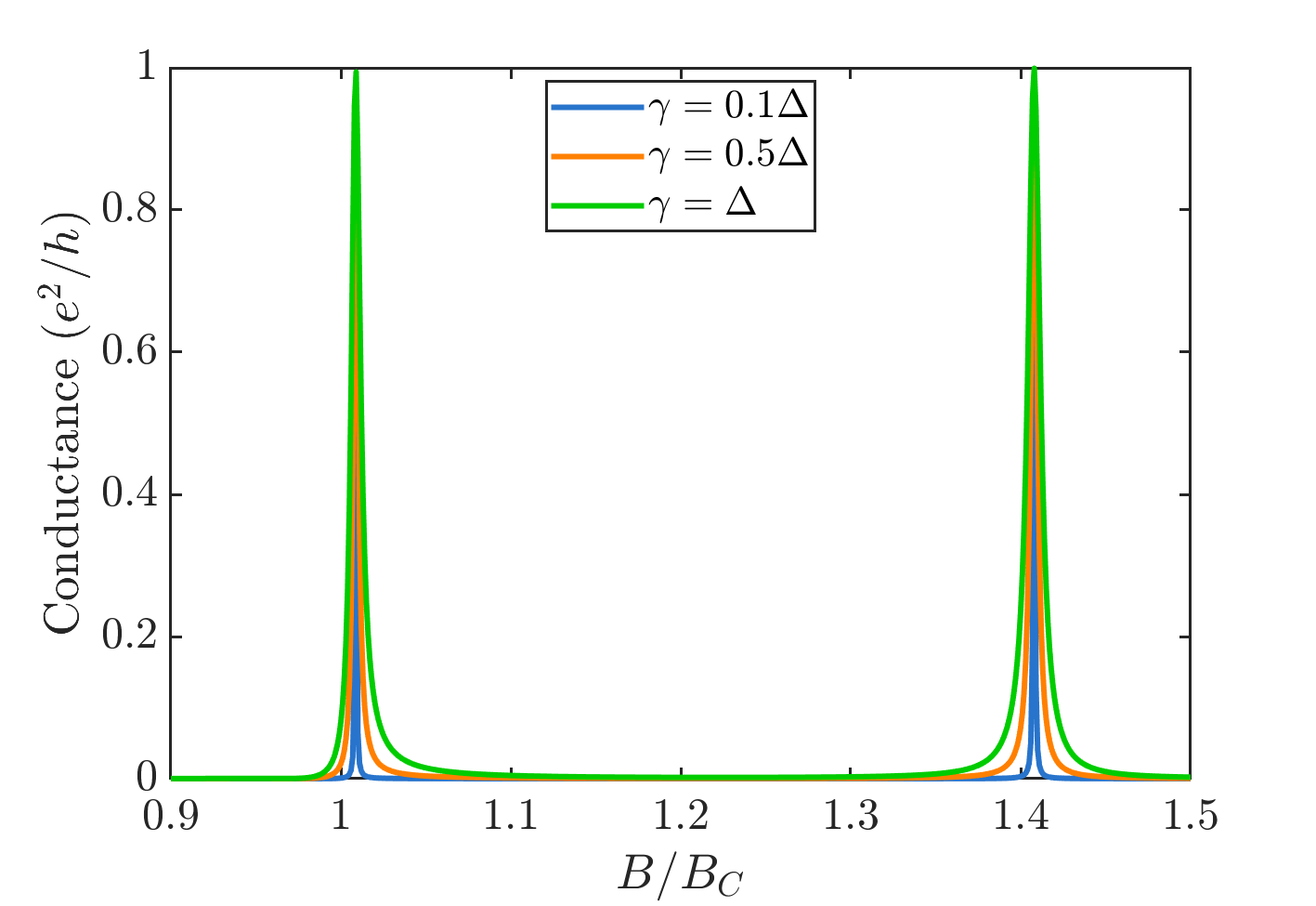}
    \label{fig:coh_condvsB}}
\caption{Coherent transport through a clean Majorana nanowire, exhibiting the quantized zero-bias conductance peak signature of zero-modes. (a) Differential conductance as a function of voltage, at magnetic field $B_1$ corresponding to the first zero-mode shows the zero-bias conductance peak quantized at $e^2/h$. (b) Zero-bias conductance with varying magnetic field shows  quantized peaks at the values of $B$ ($B_1$ and $B_2$) where the oscillating states become MZMs.}
\label{fig:coh-cond-vsB}
\end{figure}

\subsection{\label{sec:coherent}Coherent Transport}
Next, we employ the NEGF formalism in the clean wire limit in the coherent transport regime through the nanowire. Using the spectral function defined in \eqref{eq:spec_main}, we calculate the local density of states (LDOS) at a specific site in the nanowire with varying magnetic field as shown in Fig.~\ref{fig:coherent-ldos-site}. The total density of states follows the energy spectrum in Fig.~\ref{fig:spectrum-short}, but the contributions from each site vary across the eigenstates. Figure ~\ref{fig:coherent-ldos-site5} shows the LDOS at the fifth site in the $100$-site nanowire, which is a location near the left end of the wire. We see that this site mainly contributes to the oscillating mode close to zero energy, and its contribution to the other states is small. The LDOS at the center of the nanowire (site $50$) in Fig.~\ref{fig:coherent-ldos-site50} on the other hand shows a negligible contribution to the zero-mode. In contrast, the higher energy states have nearly equal LDOS at either site, which indicates that they are not localized. \\
\indent To examine more closely the localization of states, in Fig.~\ref{fig:coherent-ldos-B} we focus on the spatial variation of the LDOS particular Zeeman energies corresponding to the point $B_1$ (Fig.~\ref{fig:coherent-ldos-B1}) and $B_2$ (Fig.~\ref{fig:coherent-ldos-B2}) as marked in Fig.~\ref{fig:spectrum-short}, in which the oscillating low-energy modes in the energy spectrum cross to give  to give exact zero-energy modes. We see clear end-localization of the zero-energy Fermionic wavefunctions, with maximum density at site 5 near the edge. We also see the finite spatial spread into the bulk of the nanowire and the finite Majorana localization length. Since $l_M$ increases with increasing Zeeman field in our sysstem, the zero-modes are seen to be more localized at the point $B_1$ than at the point $B_2$ as expected. We also note that the parity crossings have a node at the center of the nanowire at $B_2$ but a non-zero value at $B_1$, which is also seen in Fig.~\ref{fig:coherent-ldos-site50} in the density of the oscillating modes.

\begin{figure}[h!]
	\subfigure[]{
		\includegraphics[height=0.35\textwidth,width=0.45\textwidth]{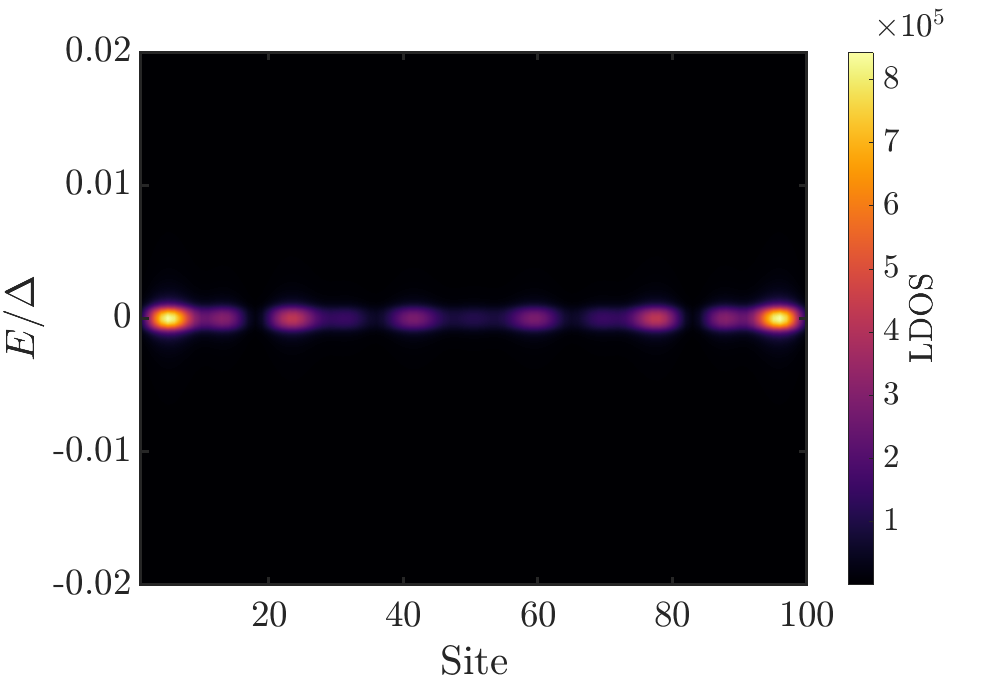}
		\label{fig:ldos-dph-d1}}
	\hfill
	\subfigure[]{
		\includegraphics[height=0.35\textwidth,width=0.45\textwidth]{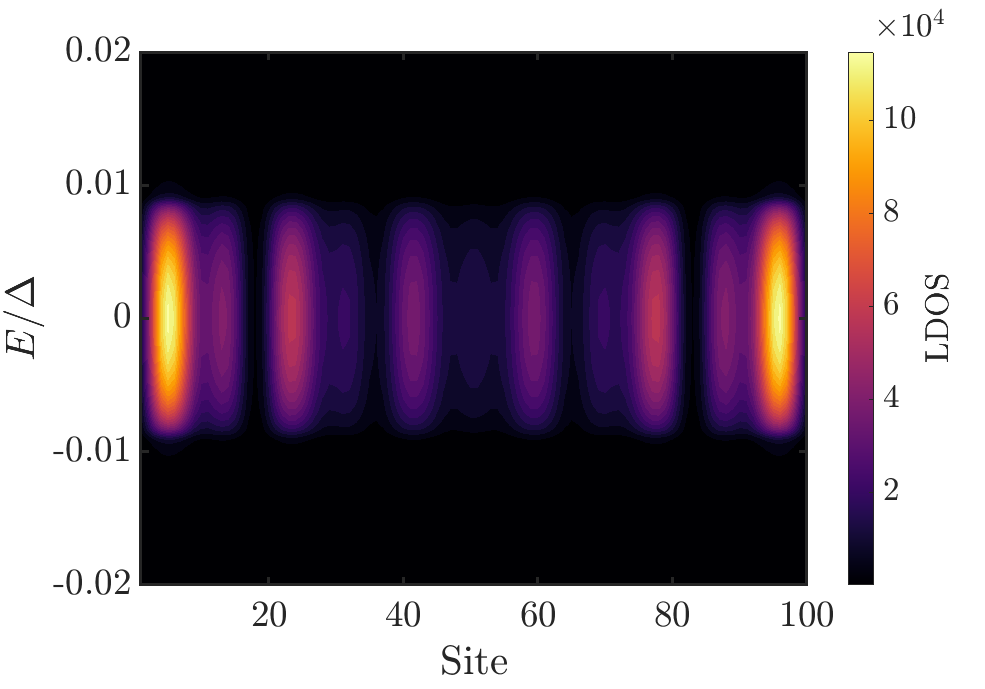}
		\label{fig:ldos-dph-d2}}
	\caption{Local density of states at $B_1$ with momentum-relaxing dephasing interactions, characterized by $d_0 = 10^{-8} t_0^2$ (a) and $d_0 = 10^{-6} t_0^2$ (b). Non-coherent effects cause the states to be broadened out in energy. However, the end-localized nature remains preserved, confirming that these modes are still topologically significant.}\label{fig:ldos-dph}
\end{figure}
\begin{figure}[h!]
	\subfigure[]{
		\includegraphics[height=0.35\textwidth,width=0.45\textwidth]{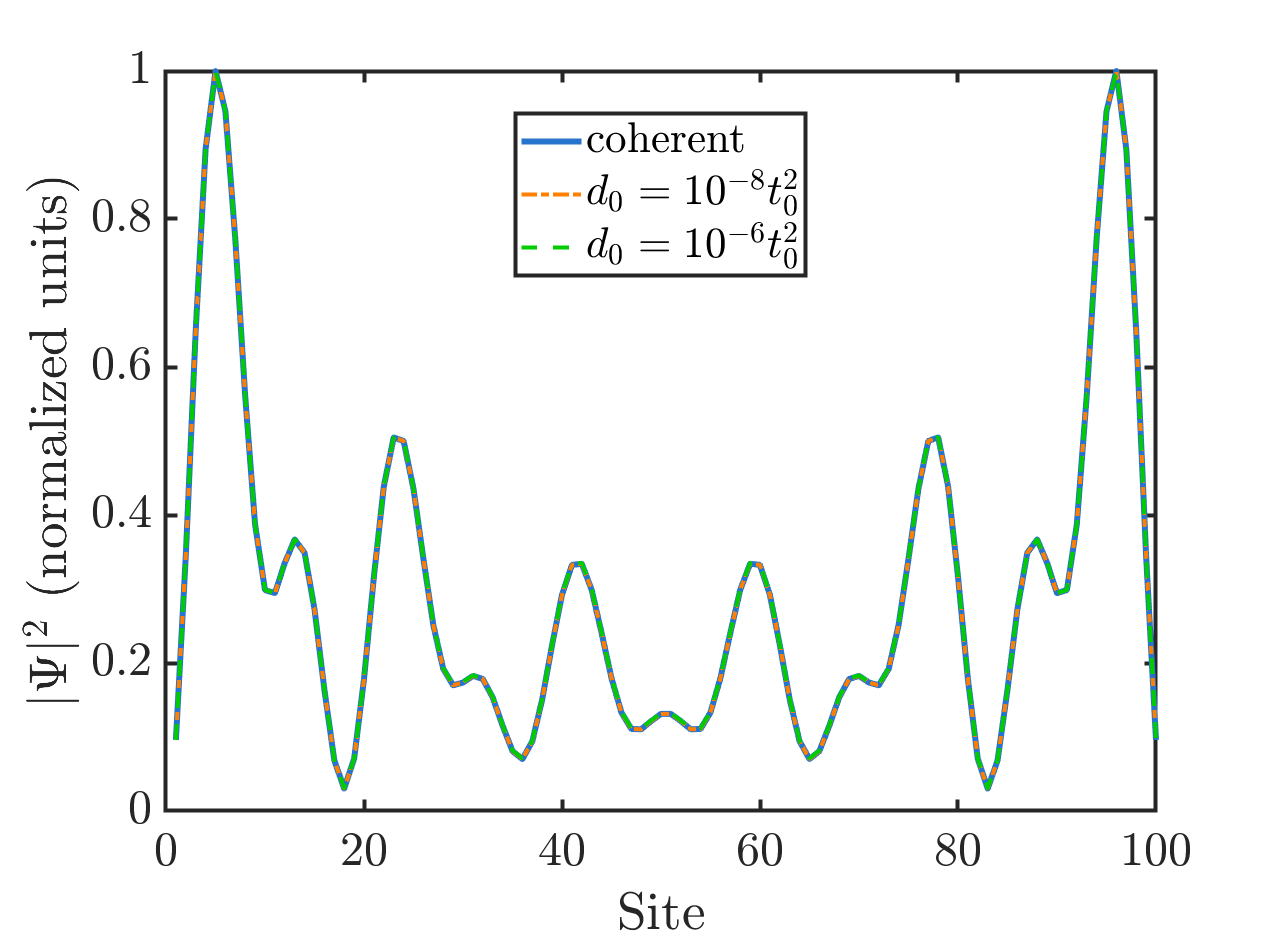}
		\label{fig:wf-B1}}
	\hfill
	\subfigure[]{
		\includegraphics[height=0.35\textwidth,width=0.45\textwidth]{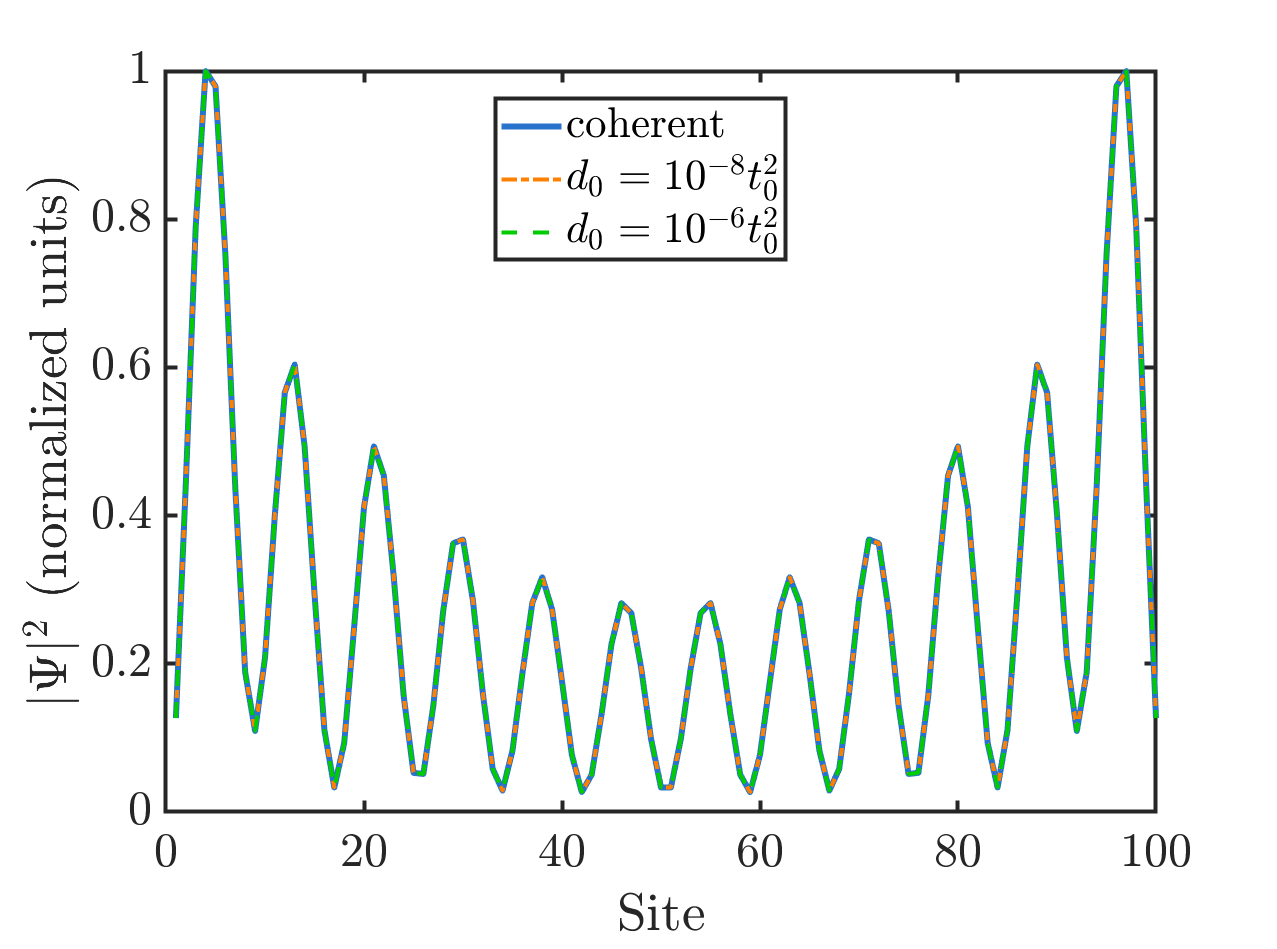}
		\label{fig:wf-B2}}
	\caption{Probability density $\mid \Psi_i \mid^2$ at each site in the nanowire at Zeeman energies $B_1$ (a) and $B_2$ (b), for a clean nanowire and for two different magnitudes of the dephasing interactions $d_0$. In all the cases, the density has exactly the same spatial variation, indicating that the localized nature of the MZMs is preserved even in the presence of fluctuating impurities.}
	\label{fig:wf}
\end{figure}
\begin{figure}[h!]
	\subfigure[]{
		\includegraphics[height=0.35\textwidth,width=0.45\textwidth]{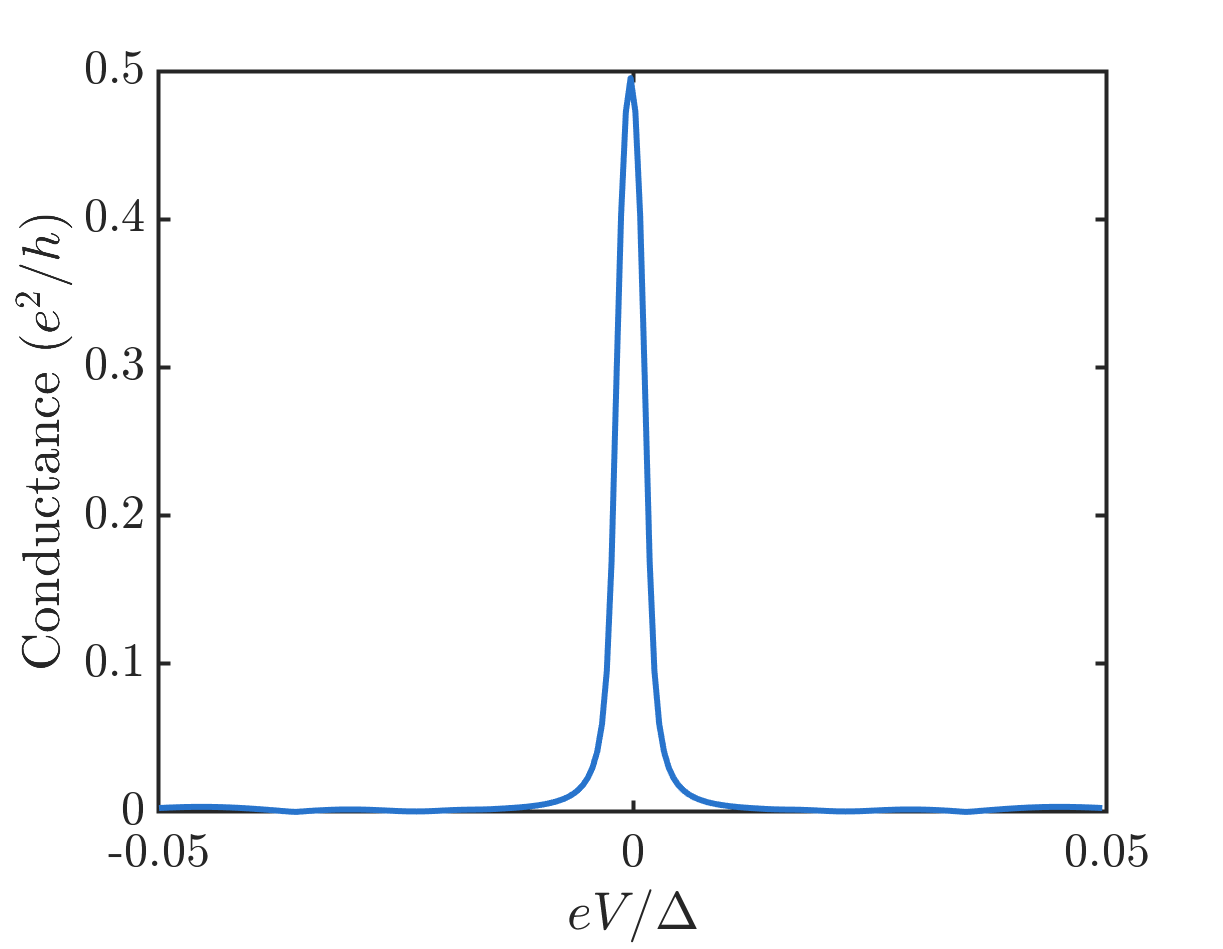}
		\label{fig:condvsV_B1_d1}}
	\hfill
	\subfigure[]{
		\includegraphics[height=0.35\textwidth,width=0.45\textwidth]{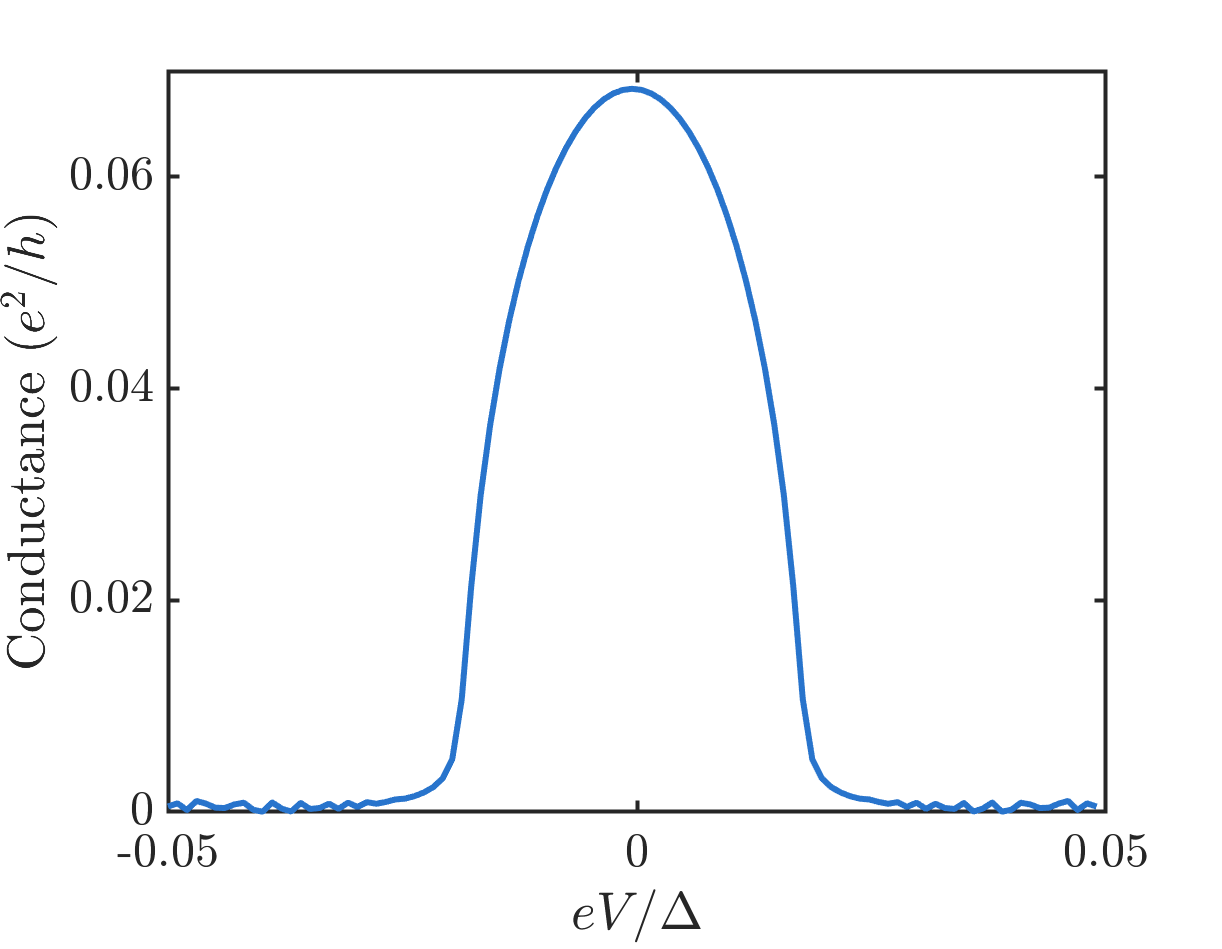}
		\label{fig:condvsV_B1_d2}}
	\caption{Differential conductance at magnetic field $B_1$, in the presence of non-coherent interactions characterized by $d_0 = 10^{-8} t_0^2$ (a) and $d_0 = 10^{-6} t_0^2$ (b). The ZBCP is present but no longer quantized at $e^2/h$, and its magnitude decreases with increasing $d_0$, along with a significantly more broadened lineshape.}
	\label{fig:condvsV_B1_d}
\end{figure}
\begin{figure}
	\subfigure[]{
		\includegraphics[height=0.35\textwidth,width=0.42\textwidth]{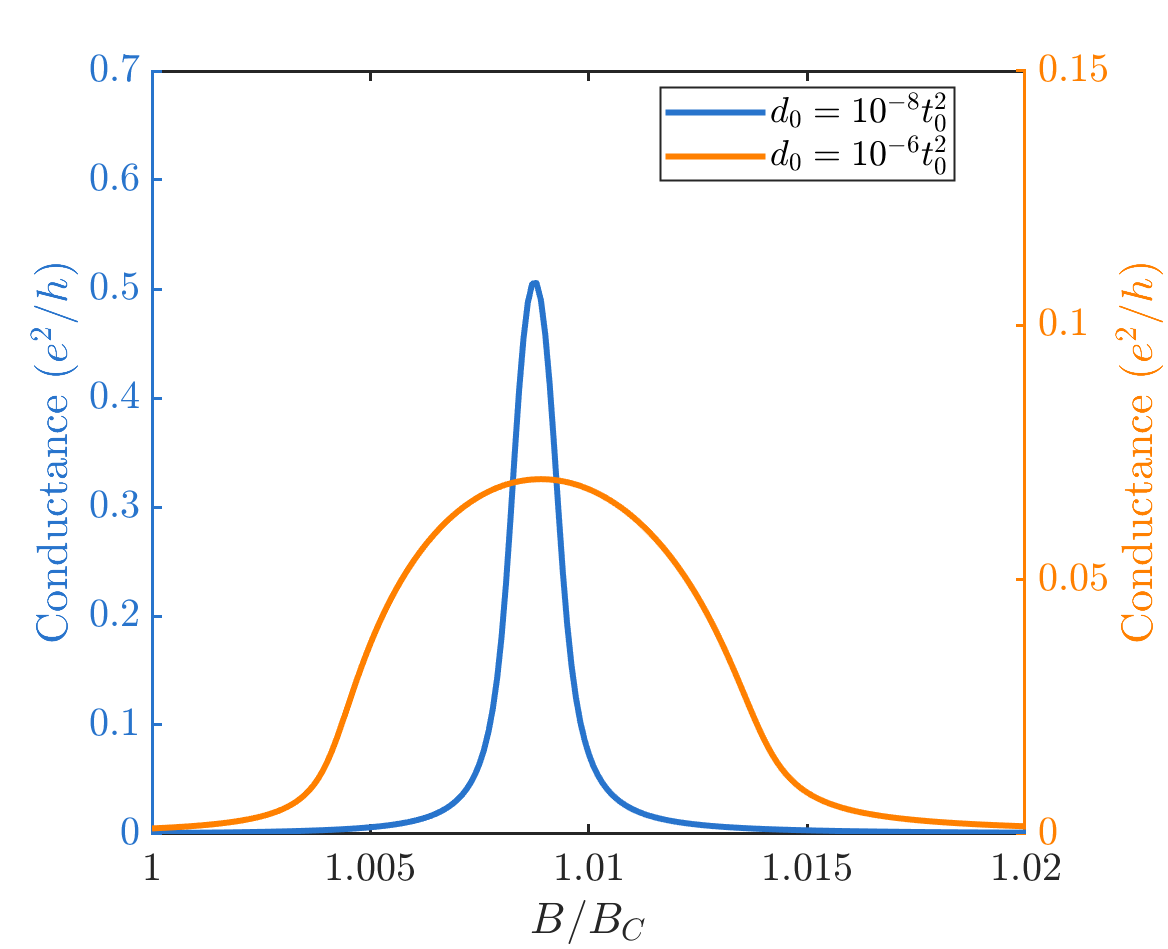}
		\label{fig:condvsB1_dph}}
		\hfill
	\subfigure[]{
		\includegraphics[height=0.35\textwidth,width=0.42\textwidth]{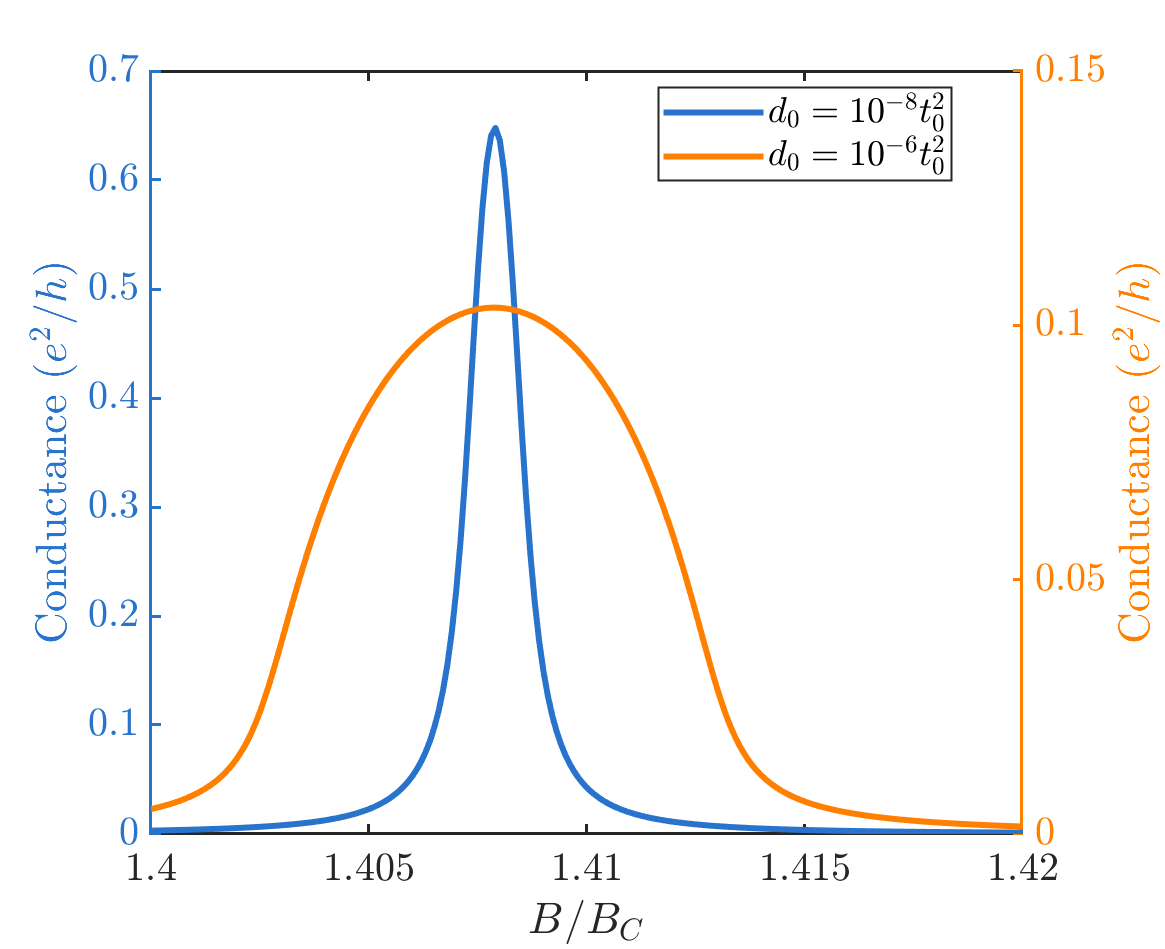}
		\label{fig:condvsB2_dph}}
	\caption{Zero-bias conductance near $B_1$ (a) and $B_2$ (b) for two different strengths of non-coherent interactions. The peaks are slightly higher at $B_2$ compared to $B_1$. The lineshapes are more broadened in B for stronger interactions.}
	\label{fig:condvsB_dph}
\end{figure}

\indent Having verified the presence of edge-localized zero-energy MZMs at specific Zeeman energies, we now turn our attention to the transmission through the nanowire. We focus on the the zero-energy transmission $T(E=0)$ near the crossing point Zeeman energies corresponding to $B_1$ and $B_2$ and for two different values of the contact coupling $\gamma$, in Fig.~\ref{fig:coherent-trans-vs-B}. The total transmission has a peak of magnitude unity at $B_1$ and $B_2$, and drops to zero at all other intermediate values of magnetic field. The total zero-energy transmission is directly related to the zero-bias conductance as noted in Sec.~\ref{sec:negf}.

\begin{figure}
\centering
\includegraphics[height=0.35\textwidth,width=0.42\textwidth]{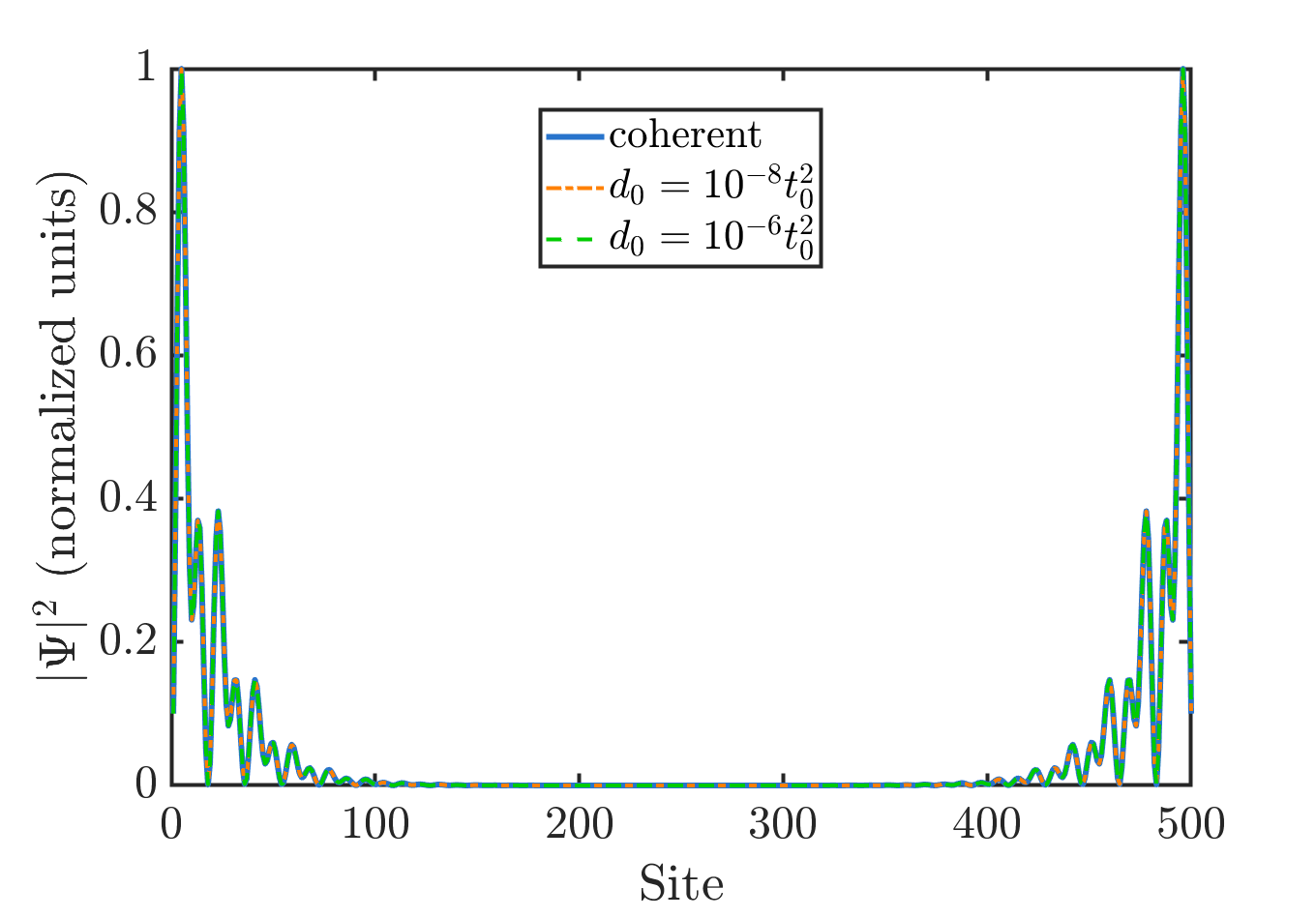}
	\caption{Probability density $\mid \Psi_i \mid^2$ at each site in the nanowire at a magnetic field $B > B_C$ in the topological regime, for a clean nanowire and for two different magnitudes of the dephasing interactions $d_0$. The end-localization is better in this case, with zero probability density for most region in the bulk of the wire. As in Fig.~\ref{fig:wf}, the spatial variation is the same in the coherent and non-coherent regimes.}
	\label{fig:long_wf_dph}
\end{figure}

\begin{figure}
\centering
\includegraphics[width=0.5\textwidth]{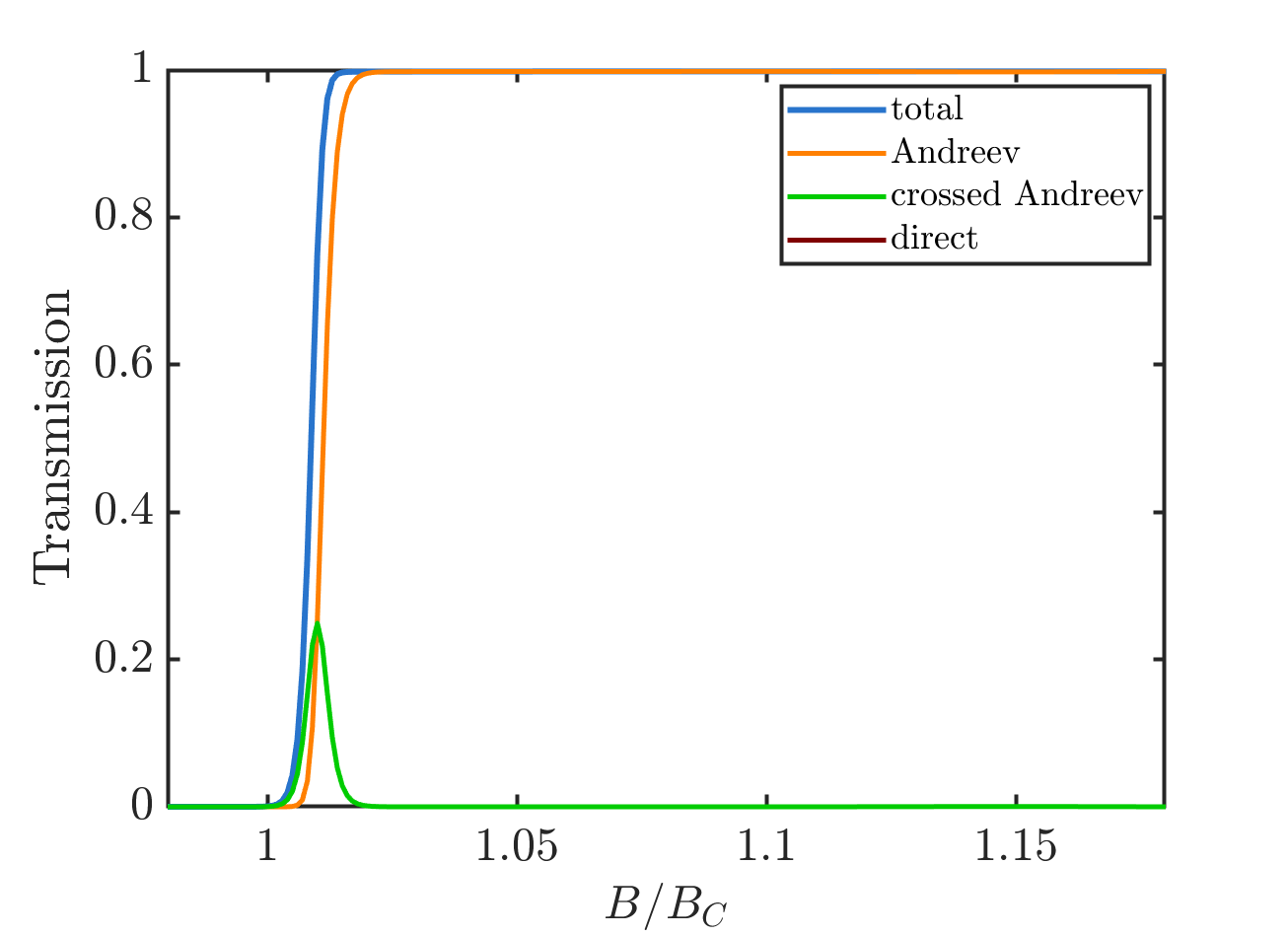}
	\caption{Local and non-local transmission at zero energy with varying magnetic field for a long wire ($l=8\mu m$), resolved into the direct, Andreev and crossed Andreev process components. The total transmission shows a plateau of magnitude 1 in the topological regime, dominated by Andreev reflection.}
	\label{fig:long_TvsB}
\end{figure}

\begin{figure}
	\subfigure[]{
		\includegraphics[height=0.35\textwidth,width=0.42\textwidth]{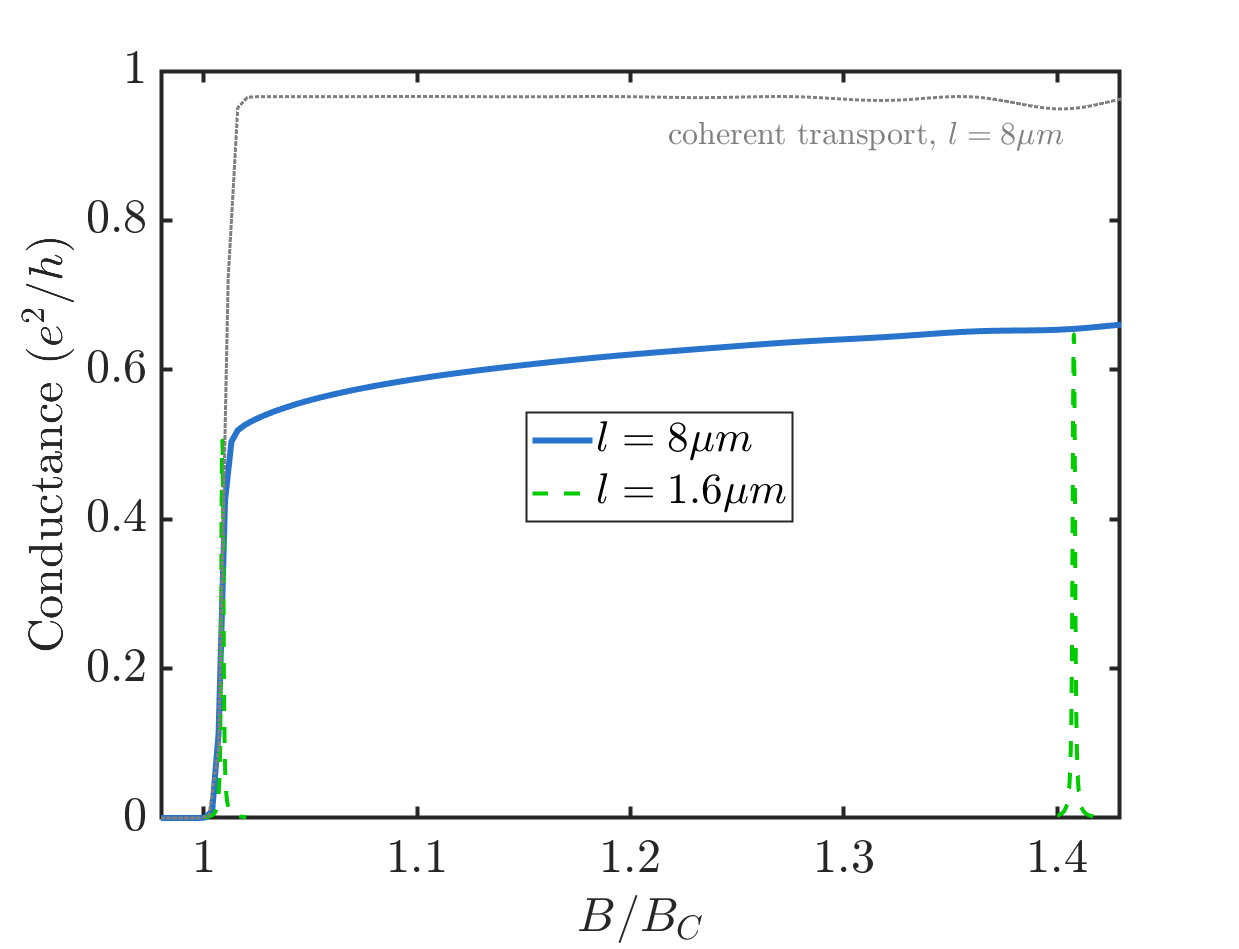}
		\label{fig:long_condvsB_d1}}
		\hfill
	\subfigure[]{
		\includegraphics[height=0.35\textwidth,width=0.42\textwidth]{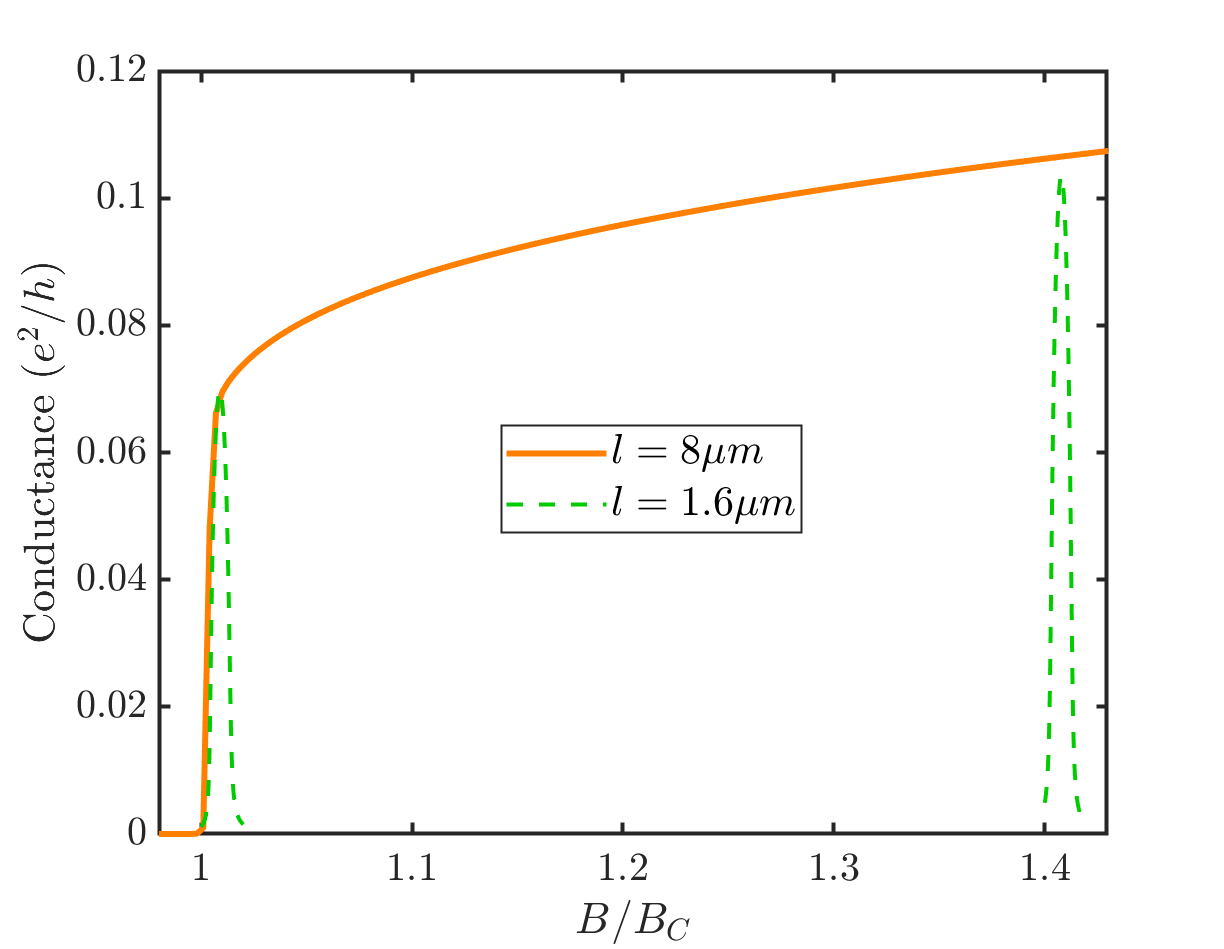}
		\label{fig:long_condvsB_d2}}
	\caption{Zero-bias conductance for a longer nanowire with varying magnetic field for two different strengths of non-coherent interactions, $d_0 = 10^{-8}t_0^2 $ in (a) and $d_0 = 10^{-6}t_0^2 $ in (b). The corresponding conductance for the short wire are shown as green dashed lines. The coherent conductance is also shown for reference in (a) as a dotted grey line.}
	\label{fig:long_condvsB_dph}
\end{figure}
\indent We note that the dominant process when the modes have exactly zero energy is the local Andreev reflection process. At Zeeman energies close to the parity crossings corresponding to $B_1$ or $B_2$, the non-local direct and crossed Andreev components have a significant non-zero contribution to transmission, but it drops to a minimum at precisely $B_1$ and $B_2$, leading to their split-peaks feature in Fig.~\ref{fig:coherent-trans-vs-B}. We also note the significant robustness of the peak height to varying contact coupling. Increasing the contact coupling causes the states to be more broadened out in energy, while maintaining the unity transmisison peak. In Figs.~\ref{fig:coherent-trans-vs-B1-gam2} and \ref{fig:coherent-trans-vs-B2-gam2}, the contact coupling $\gamma$ is ten times that in Figs.~\ref{fig:coherent-trans-vs-B1} and \ref{fig:coherent-trans-vs-B2}, which manifests as broadening of the transmission peak. The relative contributions of the three processes may change but the peak height of the total transmission is preserved at unity. It is also worth noting by comparing Fig.~\ref{fig:coherent-trans-vs-B1} and  Fig.~\ref{fig:coherent-trans-vs-B1-gam2} with  Fig.~\ref{fig:coherent-trans-vs-B2} and Fig.~\ref{fig:coherent-trans-vs-B2-gam2} that the individual contributions of the local and non-local processes change at the two unique parity crossings, while the total magnitude remains quantized at unity at the magnetic field corresponding to either crossing. This point will be taken up in detail in the discussion section and is intimately related to the localization length and the competition between the direct and crossed Andreev processes. In fact, the broadening with respect to magnetic field leads to a range of $B$ at which the transmission is nearly unity. This is one possible mechanism that would lead to robustness to varying Zeeman field, as is observed in experiments.\\
\indent We now analyze quantum transport through the N-TS-N system, as shown in Fig.~\ref{fig:schematic}, by applying a symmetric bias. The current in the NEGF formalism is given by Eqs.~\ref{direct_main}, \ref{Andreev_main} and \ref{Crossed_Andreev_main}. It is worth noting that, due to the symmetric biasing,  $f_L^{ee} = f_R^{hh}$, and hence, the crossed Andreev transmission processes does not contribute to the current \cite{leumer2020linear}, and the non-local current only consists of direct transmission contribution. Also, for the N-TS-N three terminal set up, the conductance is quantized at $e^2/h$ instead of $2e^2/h$ due to the presence of two N-TS junctions in series and the resulting voltage divider \cite{leumer2020linear,Luting}, as opposed to only one junction in a typical two terminal N-TS set up. \\
\indent We first focus on the conductance at the magnetic field corresponding to $B_1$ in Fig.~\ref{fig:coherent-cond-vs-V}. As expected, we observe a zero-bias conductance peak of magnitude $e^2/h$. The peak height remains robust to varying contact coupling $\gamma$ and the peak is seen to be broadened in the voltage axis. We reiterate that the peak is expected to be quantized at $e^2/h$ for symmetric biasing, unlike the experimentally measured value of $2e^2/h$, which corresponds to asymmetric biasing $V_L = V$, $V_R = 0$ \cite{leumer2020linear,Luting}. \\
\indent Next, we vary the magnetic field $B$ and look at the ZBCP by applying a small voltage ($V \approx 10^{-5} \Delta$) where $G \approx I/V$. In Fig.~\ref{fig:coh-cond-vsB}, we observe quantized conductance peaks at $B_1$ and $B_2$, with the quantization robust to changing contact coupling. As seen in the transmission components, the ZBCP is dominated by the local Andreev reflection process.
\subsection{\label{sec:non-coherent}Effect of dephasing}
We now turn the attention to the effect of dephasing to demonstrate its effect using two values of the non-coherent interactions $d_0 = 10^{-8}t_0^2$ and $d_0 = 10^{-6}t_0^2$. 
\indent We calculate the LDOS at each site as a function of energy, at the magnetic field corresponding to $B_1$ at the two values of interaction strength in Fig.~\ref{fig:ldos-dph}. We see that the states are broadened out in energy and this broadening increases with the strength of interactions $d_0$. However, the zero-energy states remain localized at the edges with maximum density at the fifth site, as before. This points to the preservation of how local the MZM is even in the presence of non-coherent interactions, that is, there is no spatial broadening.\\
\indent To reinforce the nature of localization and the changes due to non-coherent processes, we compute the probability density of these states in the nanowire, at site $i$ as
\begin{equation}
\mid \Psi_i \mid^2 = \int -i G^{<}(i,i) dE. 
\end{equation}
Here we note that the integral must run over the entire energy range over which the MZM is broadened. Once that is ensured, we surprisingly see no change in the spatial variation of the probability density $\mid \Psi_i \mid^2$ as we go from a clean nanowire to stronger non-coherent interactions in Fig.~\ref{fig:wf}. This is a strong indicator that the edge-local nature of the zero-modes remains completely unchanged. Further, the wavefunctions match the localization seen in the LDOS, with the fifth and the ninety fifth sites having the highest density of states. The oscillations in density due to hybridization of the end-localized states are also seen clearly here.\\
\indent Finally, we calculate the transport through the nanowire with non-coherent interactions using the current formula in \eqref{Full_Iop_Nambu}. We reiterate that there is no simple relation between current and transmission as in \eqref{direct_main}, \eqref{Andreev_main} and \eqref{Crossed_Andreev_main}, and hence cannot express the currents in terms of the local and non-local contributions like how we could in the coherent case. \\
\indent As seen in Fig.~\ref{fig:condvsV_B1_d}, the differential conductance thus calculated still exhibits a ZBCP at the parity crossings, even in the presence of dephasing and momentum-relaxation. However, we observe a marked decrease in the height of this ZBCP - and that it is no longer quantized, unlike in the case of increasing contact broadening, which preserved the quantization. 
The peak is further suppressed for stronger non-coherent interactions, dropping by almost one order of magnitude in Fig.~\ref{fig:condvsV_B1_d2}. As expected, the ZBCP shows significant broadening in voltage, corresponding to the energy-broadening of the states.\\
\indent Next, we vary the magnetic field $B$ and examine the peak value of the conductance in Fig.~\ref{fig:condvsB_dph}. We now note that the conductance is significantly broadened along the $B$ axis as well, for higher values of the interaction strength $d_0$. Notably, the peak height is higher at $B_2$ (Fig.~\ref{fig:condvsB2_dph}) as compared to $B_1$ (Fig.~\ref{fig:condvsB1_dph}), which points to a possible variation in the robustness of the ZBCP at different parity crossings.\\
\subsection{\label{sec:long_wire}Long Nanowire}
Having shown some key features related to the Majorana ZBCP in shorter wires, in the coherent and non-coherent cases, specifically at the parity crossings, we now take a quick look at the case of the long wire by considering the case depicted in Fig.~\ref{fig:spectrum-long} where parity crossings cannot be discerned as easily. 
The inclusion of dephasing and momentum-relaxation as done earlier in Sec.~\ref{sec:non-coherent} preserves the end-localization as seen in the energy-integrated probability densities as seen in Fig.~\ref{fig:long_wf_dph}. The densities when integrated over the energy range over which the MZMs are broadened, remain the same as in the clean nanowire limit in the presence of non-coherent interactions of any strength. Since the wire length is much greater than the Majorana localization length in this case, we see a strong localization of the Majorana modes to the ends of the nanowire, with the probability densities going to zero for a large part of the bulk region in the middle. \\ 
\indent  In the long wire, the amplitude of Majorana oscillations in the energy spectrum is smaller than the thermal and contact broadening of the energy states. Therefore, we see unity transmission at magnetic fields above the critical field, i.e., $B > B_C$, clearly seen in Fig.~\ref{fig:long_TvsB} arising from perfect local Andreev reflection. We also note (not shown here) that discrete parity crossings indeed appear at much larger magnetic fields and can be noted via an oscillatory transmission profile. Further, in Fig.~\ref{fig:long_TvsB}, we also note that the total transmission follows the local Andreev transmission with a minor non-local contribution right before the onset of the zero bias plateau, which is related to the small contribution the non-local processes make right before the critical field is reached. This leads to the ZBCP quantized at $e^2/h$ for all $B > B_C$ for a clean wire being a plateau instead of discrete peaks, as seen in the grey dotted line in Fig.~\ref{fig:long_condvsB_d1}. \\
\indent Finally, we look at the conductance in the non-coherent regime. A ZBCP is present as for the short wire, but the peak height gets suppressed as the strength of non-coherent interactions increases. This effect is best seen in Fig.~\ref{fig:long_condvsB_dph} in the zero-bias conductance with varying magnetic field $B$. As expected, for the long wire, we have a conductance plateau instead of discrete peaks at parity crossings as was seen in the case of the short wire. However the value of the zero-bias conductance (ie., the peak height) remains almost the same as that at the crossing Zeeman energies corresponding to $B_1$ and $B_2$ of the short wire. Thus, we note that the zero-bias peak is no more robust to dephasing in a longer nanowire than in a short one, which is a very interesting result indicating a lack of length dependence of the MBS conductance even as we transition to the diffusive limit. Besides this, we also note that the magnitude of the conductance increases as the B-field increases, indicating to some extent the greater degree of robustness to dephasing. 
\section{\label{sec:discussions}Discussion}
There are several aspects in this paper that merit further consideration. Figure~\ref{fig:coherent-trans-vs-B} requires the first discussion that summarizes some important aspects of coherent transport across an MZM and the related ZBCP signatures in a three terminal set up. We note that for the second parity crossing, comparing Fig.~\ref{fig:coherent-trans-vs-B1} and  Fig.~\ref{fig:coherent-trans-vs-B1-gam2} with  Fig.~\ref{fig:coherent-trans-vs-B2} and Fig.~\ref{fig:coherent-trans-vs-B2-gam2}, that the individual contributions of the local and non-local processes change at the two unique parity crossings, while the total magnitude remains quantized at unity at the magnetic field corresponding to either crossing. Most importantly, the direct transmission process also contributes to the total conductance with a corresponding reduction in the local Andreev reflection. This can be attributed to the fact that for the realistic parameters used in our paper, the Majorana localization length actually increases between the two parity crossings \cite{cayao2017hybrid}. For the short wire, longer localization lengths lead to overlaps due to which the non-local transmissions will increase. Furthermore, the direct transmission dominates over the crossed Andreev process due to the fact that the hopping parameter is much larger than the superconducting pairing parameter. Note that there is an effective p-wave pairing that results in a competition between the direct transmission process and the crossed Andreev transmission. In the case of the long wire, it is clear from Fig.~\ref{fig:long_wf_dph}, that the length of the wire is much greater than the localization lengths, leading one to note that the only dominant process above the critical field is the local Andreev reflection process.\\
\indent In the non-coherent limit, the values of dephasing introduced can be mapped to the elastic momentum relaxation lengths, that are in turn dependent on the magnetic field. The first important aspect that arises out of dephasing is the broadenend line shape of the ZBCP as witnessed in Fig.~\ref{fig:condvsV_B1_d}. Specific comparison between Fig.~\ref{fig:condvsV_B1_d} (a) and Fig. ~\ref{fig:condvsV_B1_d}(b) clearly depicts a drastic reduction in the peak height along with significant broadening as the strength of the dephasing is increased. While the elastic momentum relaxation length decreases as a function of the dephasing strength $d_0$, the suppressed and broadened ZBCP is observed in 
Fig.~\ref{fig:long_condvsB_dph} to be independent of the nanowire length. This is in stark contrast to diffusive transport through the nanowire where the conductance scales as $ \approx 1/L$.\\
\indent The most important aspect here is the gradual transition from the coherent ballistic limit to the diffusive regime as the parameter $d_0$ is increased. This helps us uncover aspects of transport in such an intermediate regime where scattering events compete with coherent reflections. This aspect is well noted and summarized in Fig.~\ref{fig:long_condvsB_dph} (a) and (b), for both the long and the short wire. Firstly, we do note the interesting aspect, as mentioned before that the long wire is more or less as robust as the shorter wire when it comes to the conductance peak heights at the parity crossings. The absence of any such strong length dependence at the zero mode parity crossings indicates that the local Andreev reflection process is a dominant contributor to the total zero mode conductance, whose reduction in magnitude can be attributed to the energy broadening of the mode depicted in the LDOS plots in Fig. ~\ref{fig:ldos-dph}.  Interestingly, as pointed out before, the momentum relaxation processes contribute only to the energy broadening of the zero modes and not to the spatial broadening. 
\section{\label{sec:conclusions}Conclusions} To summarize, the nature of the conductance signatures that signals the presence of Majorana zero modes in a three terminal nanowire-topological superconductor hybrid system was analyzed in detail, in both the coherent and the non-coherent transport regimes. In the coherent regime, we pointed out contributions of the local Andreev reflection and the non-local transmissions toward the conductance signatures. In the non-coherent regime, via the inclusion of dephasing due to the fluctuating impurities and the resulting momentum randomization processes, it was pointed out that while the Majorana character of the mode is unchanged, a reduction in the conductance magnitude that scales with the strength of the impurity potentials is seen. Important distinctions between dephasing processes in the non-coherent regime and the contact induced tunnel broadenings on the conductance lineshapes were clearly elucidated.  Most importantly our results revealed that the addition of dephasing in the set up the does not lead to any notable length dependence to the conductance of the zero modes, contrary to what one would expect in a gradual transition to the diffusive limit as a result of dephasing. This work paves the way for systematic introduction of scattering processes into the realistic modeling of Majorana nanowire hybrid devices. \\ \\
{\it{Acknowledgements:}} We wish to acknowledge Supriyo Datta, Milena Grifoni, Magdalena Marganska, Nico Leumer, Sumanta Tewari, Abhishek Sharma and Atri Dutta for stimulating discussions. The research and development work undertaken in the project under the Visvesvaraya Ph.D Scheme of the Ministry of Electronics and Information Technology (MEITY), Government of India, is implemented by Digital India Corporation (formerly Media Lab Asia). This work is also supported by the Science and Engineering Research Board (SERB), Government of India, Grant No EMR/2017/002853 and Grant No. STR/2019/000030, the Ministry of Human Resource Development (MHRD), Government of India, Grant No. STARS/APR2019/NS/226/FS under the STARS scheme. The authors BM and JB acknowledge the Shastri Indo-Canadian Insitute for the Shastri Mobility Grant 2019. 
\appendix
\section{\label{sec:NEGF-appendix}Derivation of current components in the coherent limit}
Since the leads effectively form an infinite dimensional matrix, the NEGF approach begins by partitioning the channel and the leads \cite{Datta,Jauho} and working with the channel Green's function and incorporating the leads via self-energies. The retarded channel Green's function can now be written as:
\begin{equation}
G^{r}(E) = \left [ (E+i \eta)I - H_{BdG} - \Sigma^{r}_L - \Sigma^{r}_R \right ]^{-1},
\label{Ret}
\end{equation}
where $E$ is the free variable energy, and $I$ is the identity matrix of the dimension of the Hamiltonian. The energy is indeed a Fourier transform variable of the difference time of the two time Keldysh Green's function defined on the Schwinger-Keldysh contour \cite{Jauho}. For brevity, we will use the energy variable on the left hand side, with the energy dependence of the elements of the matrix product on the right hand side implicit.

Here a straightforward calculation of the lead self energies matrix elements $\Sigma^{r}_{\alpha}(i,i) = \sigma_{\alpha \in L/R}$ read \cite{Yeyati,Jauho,Datta}
\begin{widetext}
\begin{equation}
\sigma_{\alpha \in L/R}(E) = \sum_{k} \frac{\mid \tau_{\alpha,k} \mid^2}{\left ( E-\epsilon_{k \alpha}+ i \eta \right )}= {P} \int d \epsilon_{k \alpha}\frac{\mid \tau_{ \alpha,k} \mid^2 }{\left ( E-\epsilon_{k \alpha} \right )}  - i \pi \int d \epsilon_{k \alpha} \mid \tau_{\alpha,k} \mid^2 \delta(E-\epsilon_{k \alpha}),
\end{equation}
\end{widetext}
where $P$ stands for the Cauchy principal value. This integral can be recast as $\sigma_{\alpha}=\epsilon_{\alpha}(E) - i \gamma_{\alpha}/2$, where the real part $\epsilon_{\alpha}$ is calculated from the principal value integral and the imaginary part $\gamma_{\alpha}=i \left ( \sigma^{r}_{\alpha} - \sigma^{a}_{\alpha} \right )$ represents the level broadening or decay constant.\\
\indent In this work, we will focus on including the contacts described by the eigenbasis representation or the wide band limit. Here with the assumption of a broad-band contact dispersion ($V_{k \alpha,m} $ being independent of energy), the principal value integral representing the real part of the self energy vanishes and we are just left with the broadening matrix $\Gamma_{\alpha}(E)$ such that the self energies can be represented as $\Sigma^r_{\alpha}=-i \Gamma_{\alpha}/2$, which can be treated as an input parameter. Once we have the self energies, we need to evaluate the ``lesser'' Green's function or equivalently, the correlation function given by:
\begin{equation}
G^{<}(E)=G^r(E) \left (\Sigma_L^{<}(E) + \Sigma_R^{<} (E)\right ) G^a(E) ,
\label{eq:Gn}
\end{equation}
where $G^a$ is the advanced Green's function, which is the Hermitian conjugate of the retarded Green's function calculated from \eqref{Ret}. The quantities $\Sigma^{<}_{L,R}(E)$ represent the in-scattering functions from leads $L$ and $R$ respectively evaluated as
\begin{eqnarray}
\Sigma_{\alpha}^{<}(E) &=& -\left [ \Sigma^r_{\alpha}(E)-\Sigma^{a}_{\alpha}(E) \right ]f_{\alpha}(E) \nonumber \\
&=& i \Gamma_{\alpha}(E) f_{\alpha}(E),
\label{eq:Sigma_less}
\end{eqnarray}
where $f_{\alpha}=f(E-\mu_{\alpha})$, the Fermi-Dirac distribution in either lead $\alpha = L(R)$ with its own chemical potential $\mu_{\alpha}$. In this formulation, all different components of the currents can then be deduced from the current operator formula, which in turn, can be derived from fundamental considerations \cite{Jauho} and the current operator at the left contact now reads:
\begin{eqnarray}
I^{op}_L(E)&=& \frac{e}{h} [ G^r(E) \Sigma_L^{<} - \Sigma_L^{<} G^{a}(E) \nonumber \\
&& +  G^<(E) \Sigma_L^{a}(E) -\Sigma_L^{r}(E) G^<(E)  ]
\label{Full_Iop}
\end{eqnarray}
and making some basic manipulations and taking the trace of the above equation to find the net charge current, we arrive at the celebrated current formula \cite{Meir-Wingreen-1992}:
\begin{widetext}
\begin{equation}
I_L(E)= \frac{ie}{h}Trace \left [ \Gamma_L(E) f_L(E) \left(G^r(E)-G^a(E) \right ) + \Gamma_L(E) G^<(E) \right ].
\label{eq:MW}
\end{equation}
\end{widetext}
The above formula can be further simplified by the notation introduced in \cite{Datta} with the spectral function $A(E)= i (G^r - G^a)$ and the electron correlation matrix $G^n(E) =-i G^{<}(E)$ as
\begin{equation}
I_L(E)= \frac{e}{h}Trace \left [ \Gamma_L(E) f_L(E) A(E) - \Gamma_L(E) G^n(E) \right ].
\label{eq:Datta}
\end{equation}
Furthermore, it can be easily shown that 
\begin{eqnarray}
A(E) &=& i(G^r-G^a)  \nonumber \\
&=& G^r i\left[ \left ({G^a} \right )^{-1} - \left ({G^r} \right )^{-1} \right ]G^a  \nonumber \\
&=&  G^r i \left [\Sigma^r-\Sigma^a \right ]G^a \nonumber \\
&=& G^r\Gamma G^a,
\label{eq:spec}
\end{eqnarray}
where $\Gamma(E) = \Gamma_L(E) + \Gamma_R(E)$. Using the above in conjunction with \eqref{eq:Gn} and \eqref{eq:Sigma_less}, along with the properties of the trace operation, yields the Landauer transmission formula for the current as  
\[
I_L= \int dE Trace \left [ \Gamma_L G^r \Gamma_R G^a \right ] \left (f_L(E) - f_R(E) \right ),
\]
where $f_{L(R)}(E)$ represents the Fermi-Dirac distribution in either lead.\\
\indent However, in the case involving superconductors, either functioning as a contact or as the central system or both, the net current through the contact $\alpha$ is the difference between electron and hole currents in Nambu space. We can now use the full fledged current operator in  \eqref{Full_Iop} to evaluate the current from first principles as
\begin{widetext}
\begin{equation}
I^{op}_L(E)= \frac{e}{h} \tau_z \left [ G^r(E) \Sigma_L^{<}(E) - \Sigma_L^{<}(E) G^{a}(E)+  G^<(E) \Sigma_L^{a}(E) -\Sigma_L^{r}(E) G^<(E) \right ],
\label{Full_Iop_Nambu_app}
\end{equation}
\end{widetext}
where $\tau_z =   I_{N \times N}\otimes \sigma_z$, where $\sigma_z$ is the Pauli-z matrix and $I_{N \times N}$ is the identity matrix of dimension $N \times N$. The next step is to take a trace of the above equation, which in general need not take the form of \eqref{eq:MW}, typically when the contacts are superconducting due to the non-diagonal structure of $\Sigma_{L(R)}$, and may lead to erroneous results, specifically when evaluating the Josephson currents. However, in our case, as the contacts are normal, and hence $\Sigma_{L(R)}$ diagonal, 
\eqref{Full_Iop_Nambu_app} indeed takes the form of \eqref{eq:MW} with the net current becoming
\[
I_{\alpha} =  \int dE \frac{\left( I_{\alpha}^{e}(E) - I_{\alpha}^{h} (E) \right )}{2} ,
\]
where each current $I_{\alpha}^{e(h)}$ can be evaluated separately using the form in \eqref{eq:MW} or \eqref{eq:Datta}. The factor of a half comes since the original Hamiltonian been doubled in order to write it consistently in the Nambu space. It is then instructive in our context to note that various matrices defined within the approach will have a matrix structure due to the electron-hole Nambu space. In particular, the contact broadening matrices etc., can be written with a general diagonal structure (due to the diagonal structure of normal contacts in Nambu space) as $\Gamma_{\alpha} = \Gamma^{ee}_{\alpha} +\Gamma^{hh}_{\alpha}$ and $\Sigma_{\alpha}^{(r,a,<)} =\Sigma_{\alpha}^{(r,a,<, ee)} +\Sigma_{\alpha}^{(r,a,<, hh)}$, where the superscripts $ee (hh)$ represent the electron (hole) diagonal part of the self energy or broadening matrix. Following this and using the above observations on the current operator formula in \eqref{eq:Datta}, we obtain the electron (hole) current across the TS as a sum of three components:
\begin{widetext}
\begin{eqnarray}
I^{e(h)}_L(E) = \frac{e}{h} \left (Trace\left ( \Gamma_L^{ee (hh)}G^r \Gamma_R^{ee (hh)} G^a \right ) \left [ f_L^{ee (hh)}(E)-f_R^{ee (hh)}(E) \right ] \right )\\  \label{direct}
\qquad +  \frac{e}{h}\left (Trace\left ( \Gamma_L^{ee (hh)}G^r \Gamma_L^{hh (ee)} G^a \right ) \left [ f_L^{ee (hh)}(E)-f_L^{hh (ee)}(E) \right ] \right )\\ \label{Andreev}
\qquad + \frac{e}{h} \left ( Trace\left ( \Gamma_L^{ee (hh)}G^r \Gamma_R^{hh (ee)} G^a \right ) \left [ f_L^{ee (hh)}(E)-f_R^{hh (ee)}(E) \right ] \right ),  \label{Crossed_Andreev}
\end{eqnarray}
\end{widetext}
where \eqref{direct} represents the direct transmission process of either the electron or the hole, \eqref{Andreev} represents the direct Andreev transmission and \eqref{Crossed_Andreev} represents the crossed Andreev transmission. At this point it is worth noting that $f_{\alpha}^{ee}(E)  = f(E-\mu_{\alpha}), f_{\alpha}^{hh}(E)=f(E+\mu_{\alpha})$. 
The matrix structure of the broadening matrix $\Gamma$ itself is diagonal such that $\Gamma_{L}^{ee}=\Gamma_L(1,1)=\gamma$, $\Gamma_{L}^{hh}=\Gamma_L(2,2)=\gamma$, $\Gamma_{R}^{ee}=\Gamma_R(2N-1,2N-1)=\gamma$, $\Gamma_{R}^{hh}=\Gamma_L(2N,2N)=\gamma$ and zeros otherwise. The Andreev transmission can then be written as $T_A(E) = \gamma^2 \mid G^{r, eh}_{11} (E) \mid^2$ , the crossed Andreev transmission as $T_{CA}(E) = \gamma^2 \mid G^{r ,eh}_{1N} (E) \mid^2$ and the direct transmission can be written as $T_{D}(E) = \gamma^2 \mid G^{r ,ee}_{1N} (E) \mid^2$. Here $G^{r, ee (hh)}_{ij}$ represents the $i,j$ element of the electron (hole) diagonal block of the retarded Green function in Nambu space and  $G^{r ,eh(he)}_{ij}$ represents that of the off-diagonal block of the retarded Green function in Nambu space.

\bibliography{main}

\end{document}